\documentclass[twocolumn,twocolappendix]{aastex63}
\usepackage{amssymb, amsfonts, amsmath,framed}

\usepackage{gensymb}
\usepackage{bm}
\usepackage{multirow}
\usepackage{mathrsfs}

\usepackage{graphicx,verbatim}
\usepackage{stackengine}
\usepackage[caption=false]{subfig}
\usepackage{threeparttable}
\usepackage{tabularx}

\expandafter\ifx\csname package@font\endcsname\relax\else
 \expandafter\expandafter
 \expandafter\usepackage
 \expandafter\expandafter
 \expandafter{\csname package@font\endcsname}
\fi
\expandafter\ifx\csname package@font\endcsname\relax\else
 \expandafter\expandafter
 \expandafter\usepackage
 \expandafter\expandafter
 \expandafter{\csname package@font\endcsname}
\fi
\newcommand{\figg}[1]{Fig.~\ref{fig:#1}}
\def\bi{\begin{itemize}}
\def\ei{\end{itemize}}
\def\be{\begin{eqnarray}}
\def\ee{\end{eqnarray}}
\newcommand{\eq}[1]{Eq.~(\ref{eq:#1})}

\newcommand{\ls}{\textcolor{teal}}

\newcommand{\hz}{\textcolor{orange}}
\newcommand{\comp}{c/\omega_{\rm p}}
\newcommand{\omp}{\omega_{\rm p}}

\begin{document}
\title{Fast particle acceleration in three-dimensional relativistic reconnection}

\email{zhan2966@purdue.edu \\ lsironi@astro.columbia.edu}

\author{Hao Zhang}
\affiliation{Department of Physics, Purdue University, West Lafayette, IN, 47907, USA}

\author{Lorenzo Sironi}
\affiliation{Department of Astronomy and Columbia Astrophysics Laboratory, Columbia University, New York, NY 10027, USA}

\author{Dimitrios Giannios}
\affiliation{Department of Physics, Purdue University, West Lafayette, IN, 47907, USA}

\begin{abstract}
Magnetic reconnection is invoked as one of the primary mechanisms to produce energetic particles. 
We employ large-scale three-dimensional (3D) particle-in-cell simulations of reconnection in magnetically-dominated ($\sigma=10$) pair plasmas to study the energization physics of high-energy particles. We identify a novel acceleration mechanism that only operates in 3D. For weak guide fields, 3D plasmoids / flux ropes extend along the $z$ direction of the electric current for a length comparable to their cross-sectional radius. Unlike in 2D simulations, where particles are buried in plasmoids, in 3D we find that a fraction of particles with $\gamma\gtrsim 3\sigma$ can escape from plasmoids by moving along $z$, and so they can experience the large-scale fields in the upstream region. These ``free'' particles preferentially move in $z$ along Speiser-like orbits sampling both sides of the layer, and are accelerated linearly in time --- their Lorentz factor scales as $\gamma\propto t$, in contrast to $\gamma\propto \sqrt{t}$ in 2D. The energy gain rate approaches $\sim eE_{\rm rec}c$, where $E_{\rm rec}\simeq 0.1 B_0$ is the reconnection electric field and $B_0$ the upstream magnetic field. The spectrum of free particles is hard, $dN_{\rm free}/d\gamma\propto \gamma^{-1.5}$,  contains $\sim 20\%$ of the dissipated magnetic energy independently of domain size, and extends up to a cutoff energy scaling linearly with box size. Our results demonstrate that relativistic reconnection in GRB and AGN jets may be a promising mechanism for generating ultra-high-energy cosmic rays.
\end{abstract}


\section{Introduction}

High-energy emission from pulsar wind nebulae (PWNe) and the relativistic jets of active galactic nuclei (AGNs) and gamma-ray bursts (GRBs) raises a question about the origin of the emitting particles. 
Outflows from these compact objects are believed to be dominated by Poynting flux, i.e., the magnetic energy density is greater than the plasma rest-mass energy density. In GRB and AGN jets, magnetic field lines can reverse on small scales, as a  result of the nonlinear stages of magnetohydrodynamic (MHD) instabilities \citep[][]{romanova_92,begelman_98,spruit_01,lyutikov_03,giannios_spruit_06,bottcher_19}. Alternatively, the jet can carry current sheets from its base, like in pulsar winds \citep[][]{lyubarsky_kirk_01,drenkhahn_02a,drenkhahn_02b,kirk_sk_03,giannios_uzdensky_19,cerutti_20}. In both cases, field reversals on small scales are prone to magnetic reconnection, driving heating and particle acceleration.

Magnetic reconnection, and in particular the ``relativistic'' regime where the magnetic energy dominates over the plasma rest mass energy, is now established as an efficient mechanism of particle acceleration. Three-dimensional particle-in-cell (PIC) simulations, which offer a self-consistent description of plasma kinetics, have shown that relativistic reconnection naturally produces power-law spectra of accelerated particles \citep{zenitani_08,kagan_13,guo_14,ss_14,werner_17,guo_20_b}.
The origin of the power-law particle spectrum in two-dimensional relativistic reconnection has been recently investigated by, e.g., \citet{guo_14, uzdensky_20}.
Yet basic questions, such as how particles are accelerated to high energies, the time scale of  acceleration, and whether these processes proceed up to larger (fluid) scales, remain debated. The answer to these questions is critical when evaluating the potential of relativistic reconnection for explaining high-energy astrophysical phenomena  in relativistic outflows ({e.g.,} the emission of very-high-energy photons  or the acceleration of ultra-high-energy cosmic rays, UHECRs). For instance, \citet{giannios_10} proposed that protons escaping the reconnection layer can undergo first-order Fermi acceleration due to repeated deflections by the converging reconnection upstream flows, and can reach energies up to $E\sim10^{20}$~eV in GRB and powerful AGN jets. 


In this context, it is critical to determine from first principles the acceleration rate of the highest energy particles. PIC simulations of relativistic reconnection showed that the reconnection layer fragments into a chain of plasmoids / flux ropes \citep[e.g.,][]{sironi_16}. Recent large-scale 2D PIC simulations by \citet{petropoulou_18} and \citet{hakobyan_20} suggested that the particles populating the high-energy spectral cutoff reside in a strongly magnetized ring around the plasmoid core. Their acceleration is driven by the increase in the local field strength, coupled with the conservation of the first adiabatic invariant. They also found that the high-energy spectral cutoff grows in time as $\propto \sqrt{t}$, which appears too slow to explain, e.g., UHECR acceleration. 

These conclusions may change in a 3D geometry, which would account for the finite length of plasmoids along the $z$ direction of the electric current.
In 3D, the $z$-invariance postulated by 2D simulations can be broken by the oblique tearing instability \citep[e.g.,][]{daughton_11} and the drift kink instability \citep[e.g.,][]{zenitani_07}, which may modify the 2D picture of particle energization. While in 2D particles are efficiently trapped within plasmoids, 3D simulations of non-relativistic reconnection  \citep[][]{dahlin_2017,li_2019} have shown that 
 self-generated turbulence and chaotic magnetic fields allow high-energy particles to access multiple acceleration sites within the reconnected plasma, resulting in faster acceleration rates than in 2D.

In this work, we perform 3D PIC simulations of relativistic reconnection in a magnetically-dominated electron-positron plasma, with magnetization (i.e., the ratio of magnetic energy density to plasma rest mass energy density) $\sigma=10$. Our inflow/outflow boundary conditions allow to reliably study the statistical steady state of the system, beyond the initial transient. We identify and characterize a novel acceleration mechanism, unique to 3D. 
We find that a fraction of particles with $\gamma\gtrsim 3\sigma$ can escape from plasmoids by moving along $z$ and experience the large-scale fields in the ``upstream'' region.\footnote{We point out that the mechanism discussed by \citet{li_2019} in non-relativistic reconnection relied on particles moving between multiple acceleration sites in the reconnection ``downstream'', i.e., in the post-reconnection plasma.} 
The momentum of these ``free'' particles is preferentially oriented along $z$. They undergo Speiser-like deflections by the converging upstream flows (as envisioned by \citet{giannios_10}; see also \citet{degouveia_05}), and are accelerated linearly in time --- their Lorentz factor scales as $\gamma\propto t$. The energy gain rate approaches $\sim eE_{\rm rec}c$, where $E_{\rm rec}\simeq 0.1 B_0$ is the reconnection electric field and $B_0$ the upstream magnetic field. The spectrum of free particles is hard and can be modeled as a power law $dN_{\rm free}/d\gamma\propto \gamma^{-1.5}$ --- whose slope we justify analytically ---
extending up to a cutoff energy that scales linearly with box size. We find that the free particles account for $\sim 20\%$ of the dissipated magnetic energy, independently of domain size, yet their number (as compared to the particle count in the downstream plasma) decreases with increasing box size.


The layout of this paper is as follows. In Section~\ref{2}, we describe the simulation setup  we employ. In Section~\ref{3}, we present our main results, as regard to the particle energy and momentum spectrum, the characterization of particle orbits inside and outside the reconnection layer, and the dependence on the size of the computational domain. 
In Section~\ref{4}, we draw our conclusions and discuss implications for astrophysical systems.
We argue that relativistic reconnection in GRB and AGN jets may be a promising mechanism for generating UHECRs.

\section{Simulation Setup}\label{2}
We employ 3D PIC simulations performed with the TRISTAN-MP code \citep{buneman_93, spitkovsky_05}. The magnetic field is initialized in Harris sheet configuration, with the field along $x$ reversing at $y=0$. We parameterize the field strength $B_0$ by the magnetization $\sigma = B_0^2 / 4\pi  n_0 m c^2 = \left(\omega_{\rm c} / \omega_{\rm p}\right)^2$, where $\omega_{\rm c} = e B_0 / m c$  and $\omega_{\rm p} = \sqrt{4\pi n_0 e^2 / m}$ are respectively the Larmor frequency and the plasma frequency for the cold electron-positron plasma outside the layer, with density $n_0$. The Alfv\'{e}n speed is related to the magnetization as $v_A / c = \sqrt{\sigma/\left(\sigma + 1\right)}$; we take $\sigma = 10$. In addition to the reversing field, we initialize a uniform guide field along $z$ with strength $B_g = 0.1 \,B_0$. We have also explored a case with zero guide field and found similar results (see Tab.~\ref{tab:box}). 
We resolve the plasma skin depth $c/\omp$ with $2.5$ cells, and initialize an average of one particle in each cell. We have also tested a larger value of four particles per cell, finding no significant change in  reconnection rate, maximum energy, and particle spectra (for more details, see Tab.~\ref{tab:box}). 
The numerical speed of light is 0.45 cells/timestep. We employ periodic boundary conditions in $z$, outflow boundary conditions in $x$, while along $y$ two injectors continuously introduce fresh plasma and magnetic flux into the domain \citep[for details see][]{sironi_16,sironi_beloborodov_20}. As opposed to the commonly-adopted triple-periodic boundaries, our setup allows to evolve the system to arbitrarily long times, so we can study the statistical steady state for several Alfv\'enic crossing times. 

We trigger reconnection near the center of the simulation domain (i.e., near $x=y=0$, but along the whole $z$ extent), by removing the pressure of the hot particles initialized in the current sheet, as in \citet{sironi_16}. The characteristic $x$-length of this region is defined as $\Delta_{\rm init}$. For our largest 3D simulation (see below), we choose $\Delta_{\rm init} = 500 \comp$. For smaller boxes, we have tested different values of $\Delta_{\rm init}$, finding no difference in our main results (see Tab.~\ref{tab:box} for details). 

For our reference 3D simulation, the box length in  $x$ and $z$ (respectively, $L_x$ and $L_z$) is $\simeq4000\,{\rm cells}\sim 1600\,\comp$, while the box extent along $y$ increases over time as the two injectors recede from the current sheet. 
We also present results from a set of boxes with fixed $L_x$ but various $L_z$ from $1600\,\comp$ down to $12\,\comp$, and two sets of experiments with a fixed ratio $L_x/L_z$ ($L_x/L_z = 1$ and $L_x/L_z = 2$), but different box sizes. In the following, unless otherwise indicated, we employ our reference box with $L_x=1560\,\comp$ and $L_z= 1613\,\comp$, and we define $L=1560\,\comp$ as our unit of length.

We have also performed a 2D simulation with identical physical and numerical parameters as our reference 3D run (aside from a choice of 16 particles per cell to increase particle statistics), to emphasize 3D effects.

\section{Results}\label{3}
\begin{figure}
    \includegraphics[width=\columnwidth]{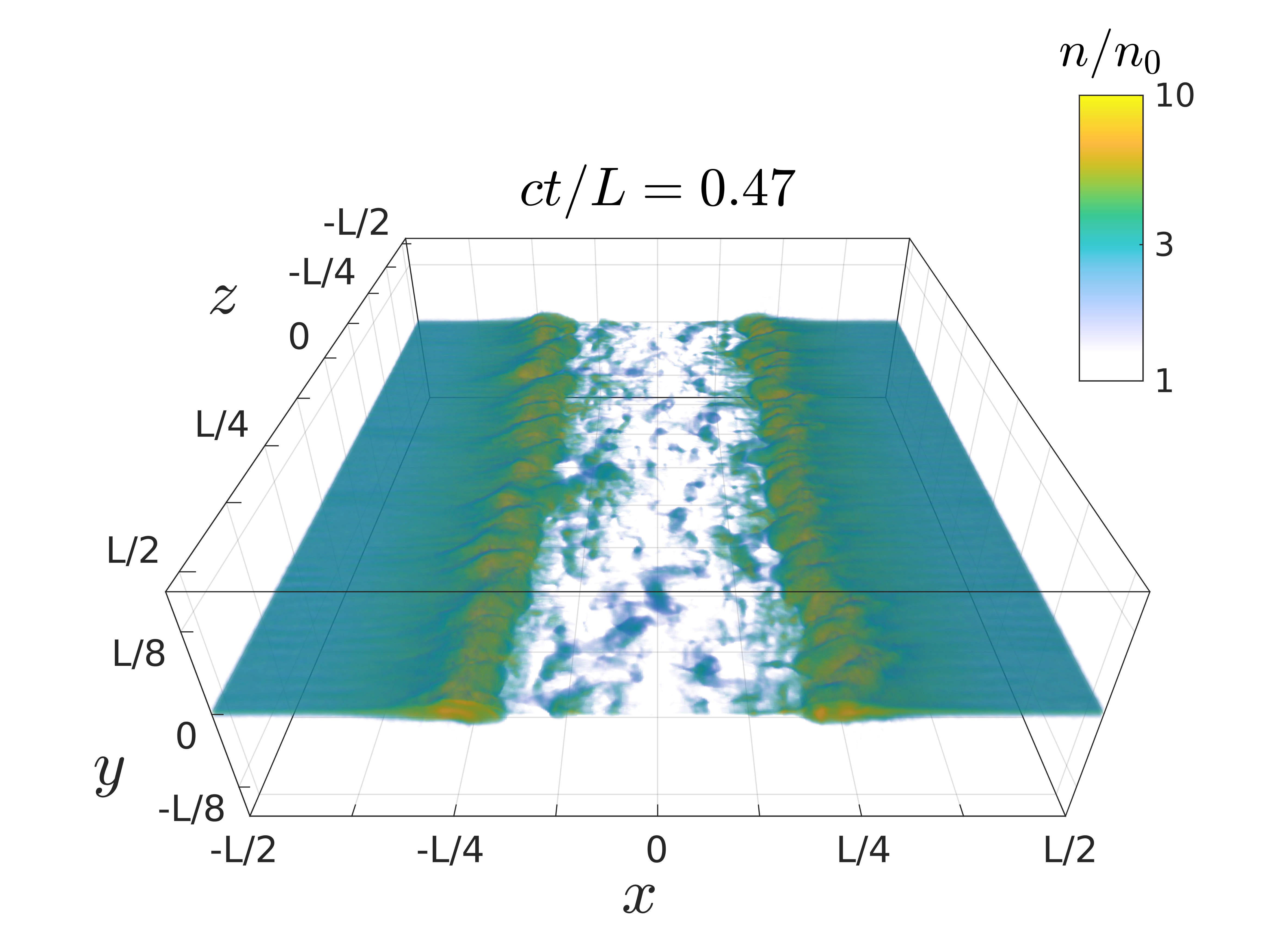}
    \includegraphics[width=\columnwidth]{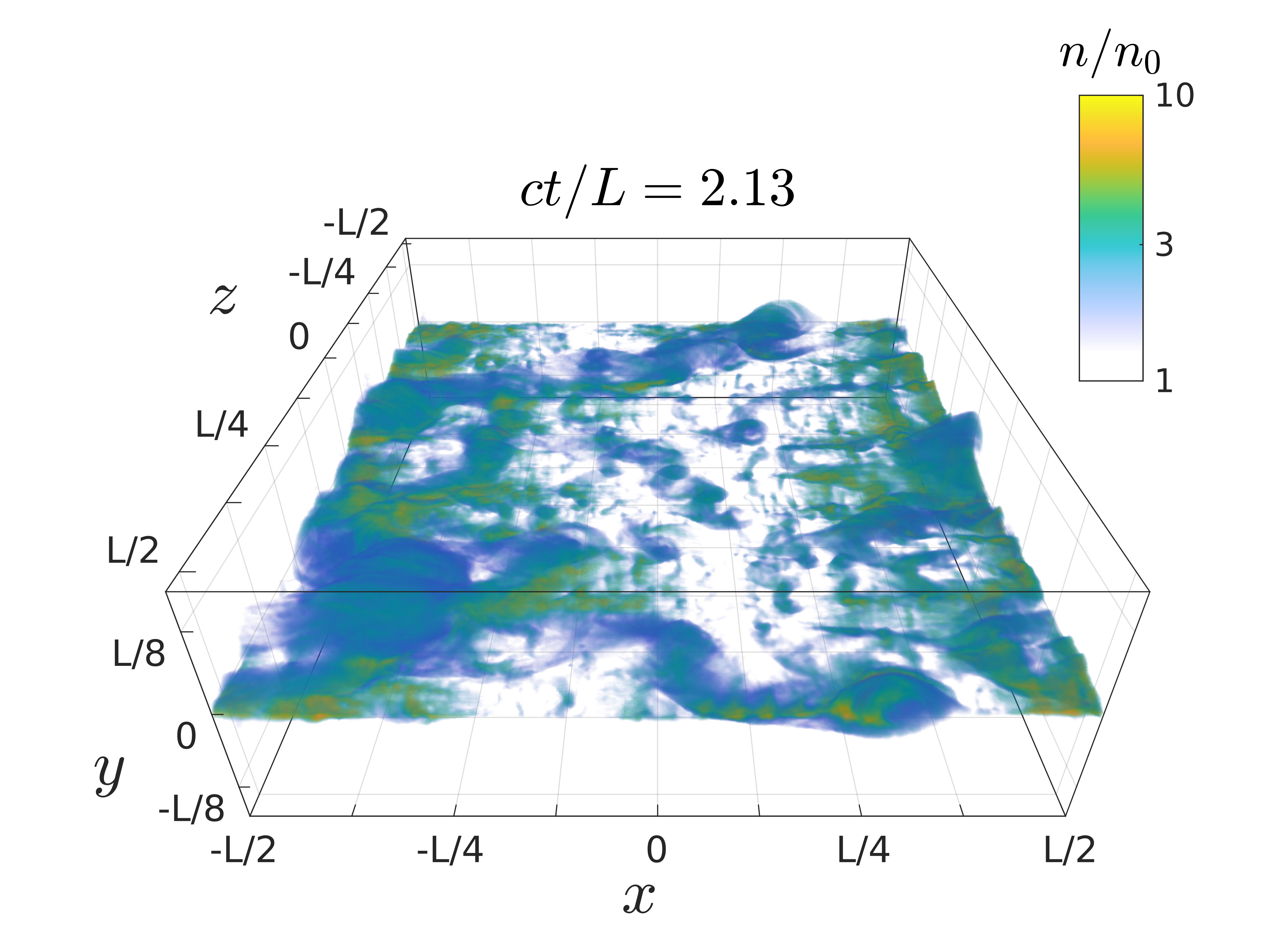}
    \caption{Two snapshots of density from our reference 3D simulation. We show the density structure at a relatively early time (top, $t = 0.47\, L/c$), when reconnection fronts are moving outwards, and at a later time (bottom, $t=2.13\, L/c$), when the system has achieved a steady state. The upstream plasma flows into the layer along $y$, while reconnection outflows move along $x$. The electric current is along the  $z$ direction, which is invariant in 2D simulations.
    }
    \label{fig:dens_3d}
\end{figure}

Fig.\ref{fig:dens_3d} shows two snapshots of the 3D density structure from our reference simulation \footnote{A Movie showing the evolution of the density structure can be found at \url{https://youtu.be/fMictkK1QNU}.}. The top panel refers to $ct/L\simeq 0.47$, and shows the two reconnection fronts (see the two overdense regions at $|x|\sim L/4$) propagating away from the center, at near the Alfv\'{e}n speed. The bottom panel of Fig.\ref{fig:dens_3d} refers to a representative time ($ct/L\simeq 2.13$) when the layer has achieved a statistical steady state. The layer is fragmented into flux ropes of various sizes, with comparable lengths in the $z$ direction as in the $x-y$ plane. The finite extent of plasmoids along the $z$ direction, likely due to the relativistic drift-kink instability  \citep{zenitani_07, zenitani_08}, plays a fundamental role for the physics of high-energy particle acceleration, as we describe below.

\begin{figure}
    \includegraphics[width=\columnwidth]{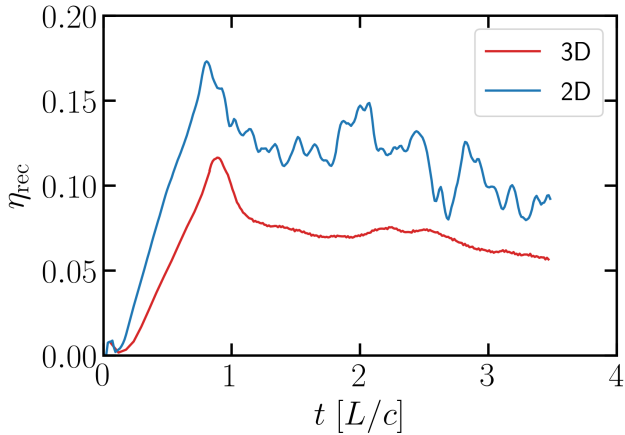}
    \caption{A comparison of the reconnection rate between 3D (red) and 2D (blue) simulations. The reconnection rate is calculated by averaging the plasma inflow velocity (in units of the speed of light) in the region $0.03L<y<0.08L$.}
    \label{fig:vin}
\end{figure}

3D instabilities can also change the reconnection rate, as compared to  2D. Fig.\ref{fig:vin} illustrates the temporal evolution of the reconnection rate $\eta_{\rm rec}\equiv v_{\rm in}/v_{\rm A}$ for both  2D (blue) and 3D (red) simulations, where $v_{\rm in}$ is the inflow speed and $v_A \simeq c$ for magnetically-dominated plasmas. The initial growth of the box-averaged reconnection rate before $ct/L\sim 0.8$ is just due to the increase of the region where reconnection is active (i.e., between the two reconnection fronts). 
When the two reconnection fronts exit the computational domain, the rate becomes quasi-steady. The reconnection rate in 3D, $\eta_{\rm rec}\sim0.075$, is slower than in 2D, $\eta_{\rm rec}\sim0.12$. 
In either case, the rate is in reasonable agreement with analytical expectations \citep{lyubarsky_05}.

The inflowing particles from the two sides of the layer mix in the reconnection region, which we shall also call ``reconnected plasma'' or ``downstream'' region. In contrast, the pre-reconnection flow shall be called ``upstream''.
To identify the region of reconnected plasma, we define a ``mixing’’ factor $\mathcal{M}$:
\begin{equation}
\mathcal{M} \equiv 1-2\left |\frac{n_{\rm top}}{n} - \frac{1}{2} \right |,
\end{equation}
where $n_{\rm top}$ is the density of particles that started from $y>0$, while $n$ is the total density. 
It follows that $\mathcal{M} = 1$ represents the  downstream plasma, where particles from the two sides of the layer are well mixed, whereas $\mathcal{M} = 0$ characterizes the upstream, where no mixing has occurred. We will use the mixing factor $\mathcal{M}$  to identify whether a particle is located in the upstream or downstream region.

\begin{figure*}
    \centering
    \subfloat{\stackunder{\includegraphics[width=0.48\linewidth]{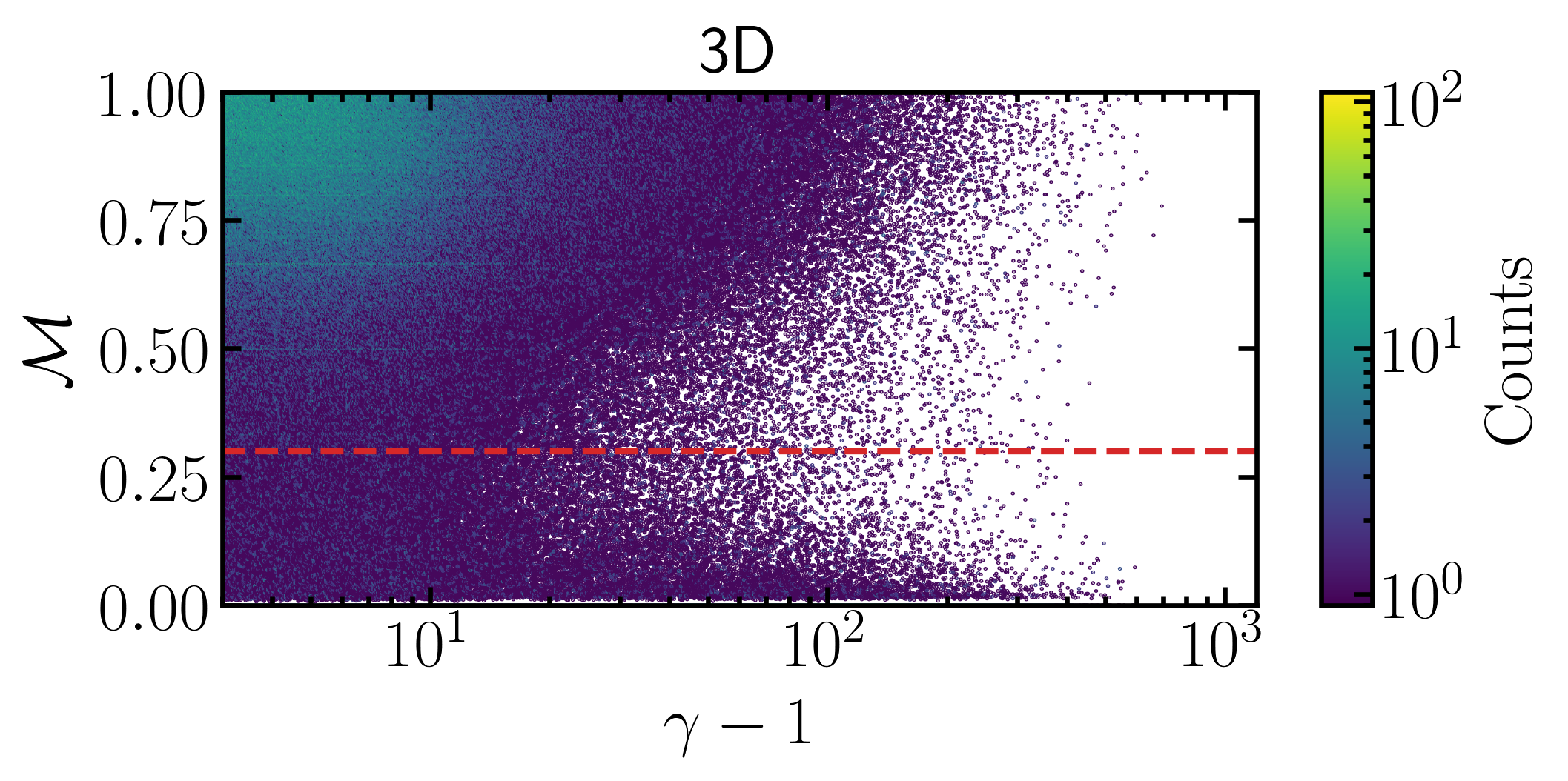}}{}}
    \qquad
    \subfloat{\stackunder{\includegraphics[width=0.48\linewidth]{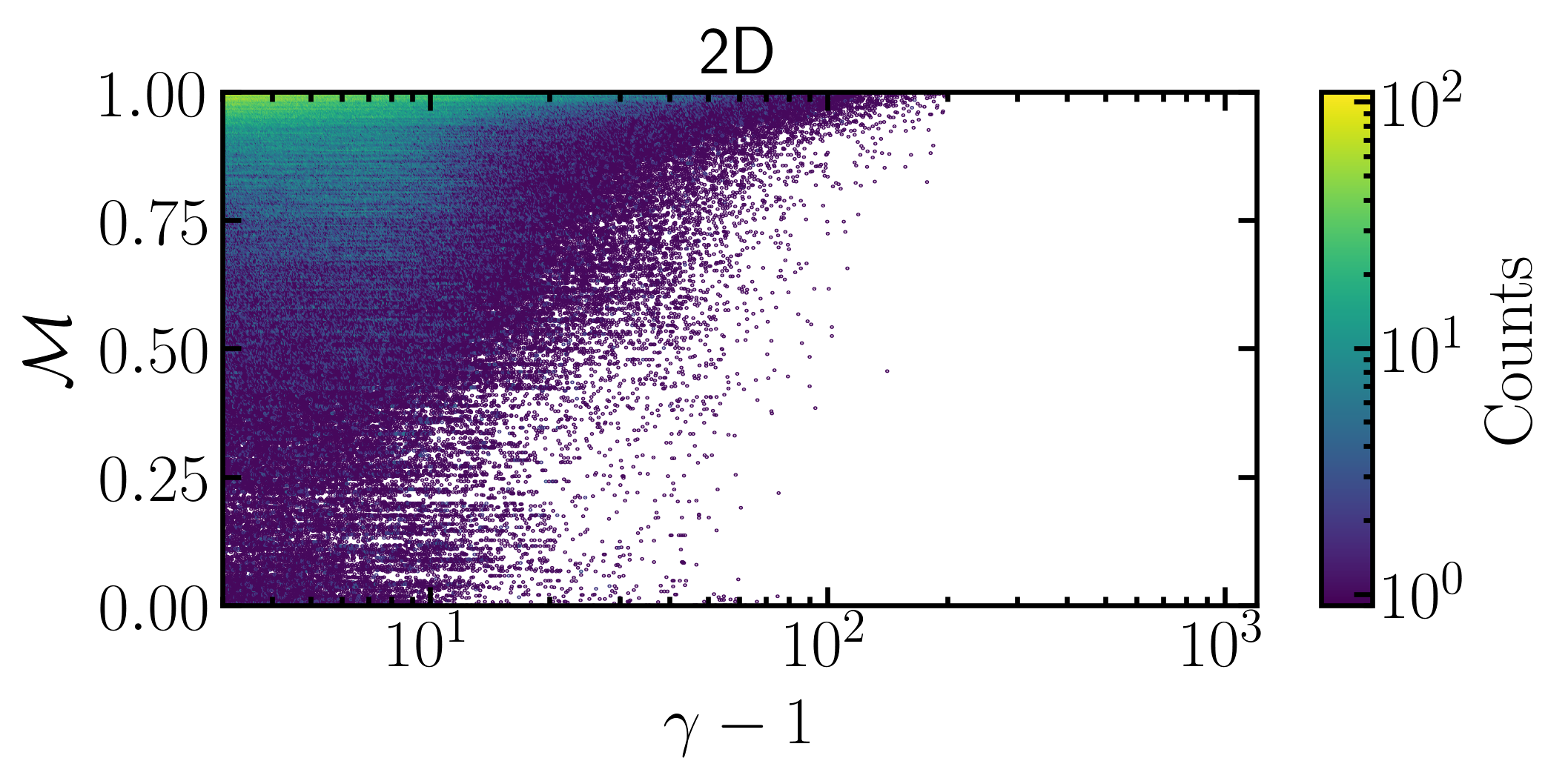}}{}}
    \caption{2D histograms of the particle Lorentz factor $\gamma$ and the mixing factor $\mathcal{M}$ (interpolated to the nearest cell) at time $t=2.37L/c$, for 3D (left) and 2D (right). The red dashed line in the left panel marks the threshold $\mathcal{M}_0=0.3$ that we employ to distinguish upstream ($\mathcal{M}<\mathcal{M}_0$) from downstream ($\mathcal{M}>\mathcal{M}_0$). }
    \label{fig:prtl_hist}
\end{figure*}

Using $\mathcal{M}$ as a criterion for separating upstream and downstream regions, we study where particles of different energies are located. Fig.~\ref{fig:prtl_hist} shows histograms of the particle Lorentz factor $\gamma$ (horizontal axis) and mixing factor $\mathcal{M}$ (vertical axis) at time $t=2.37L/c$, for 3D (left) and 2D (right) simulations. Both histograms suggest that most of the low-energy particles ($\gamma\lesssim30$) are located in the downstream region ({i.e.,} $\mathcal{M}$ near unity). In 2D, all of the high-energy particles are also located in well-mixed regions, i.e., in the downstream. In agreement with earlier studies, high-energy particles in 2D are trapped within plasmoids \citep[][]{sironi_16,petropoulou_18,hakobyan_20}.
In contrast, a significant 
fraction of high-energy particles ($\gamma\gtrsim30$) in the 3D simulation lie in low-mixing regions, i.e., in the upstream. As we show below, these are particles that have escaped from reconnection plasmoids, and are now being rapidly accelerated by the large-scale upstream fields. In the following, we will take a threshold of $\mathcal{M}_0 = 0.3$ (horizontal red dotted line in the left panel) to separate the downstream region ($\mathcal{M} > \mathcal{M}_0$) from the upstream region ($\mathcal{M} < \mathcal{M}_0$). We expect that our results will not change significantly as long as $\mathcal{M}_0$ is near $0.3$ (e.g., between 0.25 and 0.35).


In the following of this section, we first study the particle energy and momentum spectra in the 3D simulation and identify that 
high energy particles preferentially move along the $z$-direction (Section~\ref{3.1}). Then, we track particles and investigate in detail their acceleration mechanism (Section~\ref{3.2}). Finally, we investigate the dependence of our results on the domain size, in order to show that the acceleration physics should operate effectively out to larger scales (Section~\ref{3.3}).

\subsection{Particle Spectra}\label{3.1}

\begin{figure}
    \includegraphics[width=\columnwidth]{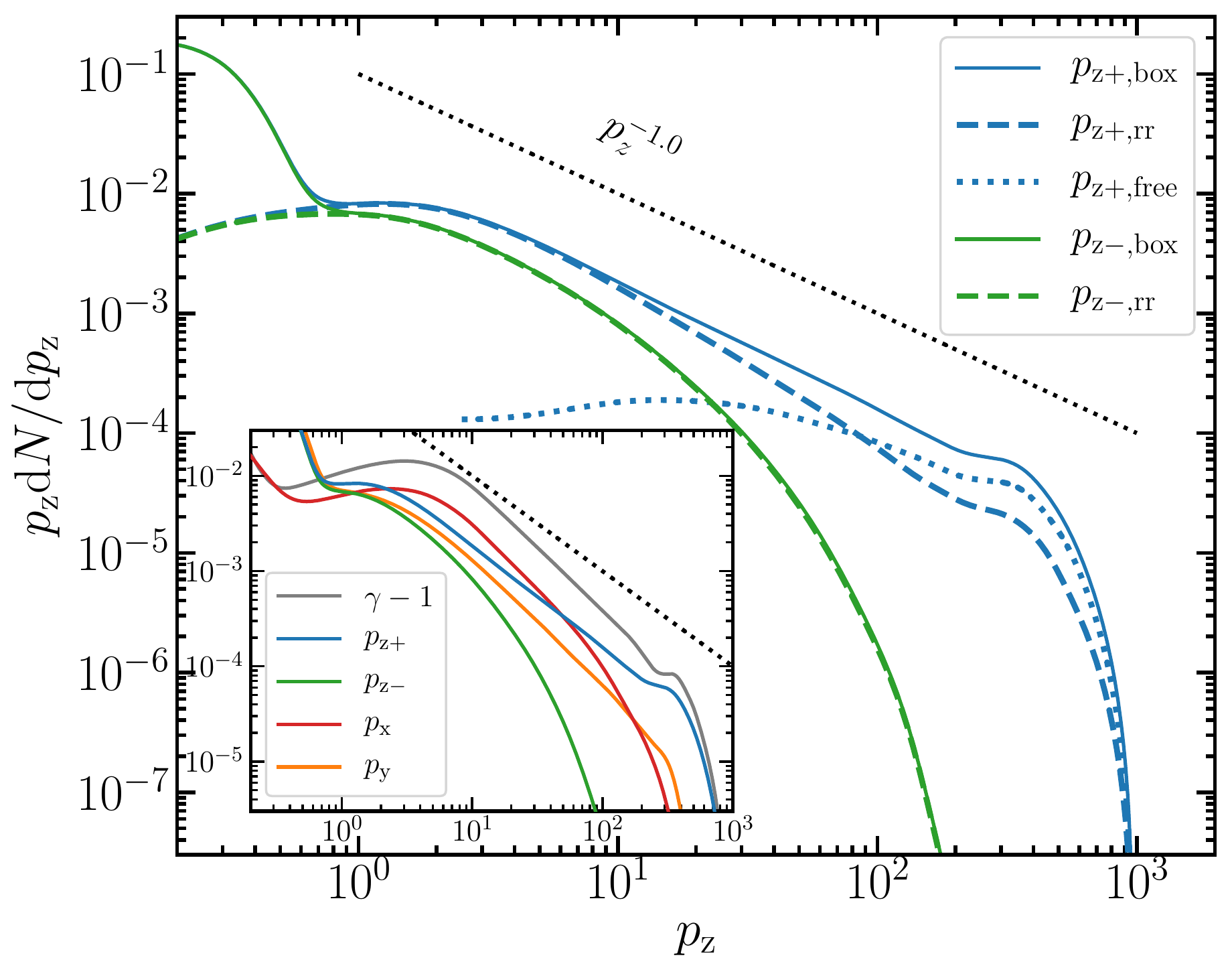}
    \caption{Momentum spectrum $p_z dN/dp_z$ of positrons, where $p_z=\gamma \beta_z$ is the dimensionless 4-velocity along the $z$ direction.  We show spectra of positrons with $p_z>0$ (blue, indicated as $p_{z+}$ in the legend) and $p_z<0$ (green, indicated as $p_{z-}$ in the legend). Spectra from the overall box are shown as solid lines (indicated with subscript ``box'' in the legend), whereas the dashed lines refer only to positrons belonging to the downstream region, as defined by the mixing condition $\mathcal{M}>\mathcal{M}_0$ (indicated with subscript ``rr'' in the legend). The spectrum of high-energy ``free'' positrons residing in the upstream region (with $\mathcal{M}<\mathcal{M}_0$), which preferentially have $p_z>0$, is indicated by the dotted blue line. The dotted black line shows a power-law $p_z^{-1}$. In the inset, we present the box-integrated positron spectra of kinetic energy (grey) and momenta in different directions, as indicated in the legend. All spectra in the main plot and in the inset are time-averaged between $t = 3.34L/c$ and $3.56L/c$ and normalized to the total number of positrons in the box.
    }
    \label{fig:spectra}
\end{figure}

A non-thermal power-law spectrum extending to high energies is a well-established outcome of relativistic reconnection \citep[e.g.,][]{ss_14}. Fig.~\ref{fig:spectra} shows the positron momentum spectrum $p_z dN/dp_z$, where $p_z=\gamma \beta_z$ is the dimensionless 4-velocity along $z$ ($\beta_z$ is the particle $z$-velocity in units of the speed of light). The spectrum is obtained by averaging between $t = 3.34L/c$ and $3.56L/c$, when the system is in steady state. The box-integrated spectrum of positrons with $p_z>0$ (blue, indicated as $p_{z+,\rm box}$ in the legend) can be modeled for $p_z\gtrsim 3$  as a power law $p_z {\rm d}N/{\rm d}p_z \propto p_z ^{-1}$.

The figure compares the momentum spectrum between positrons with $p_z>0$ (blue lines, indicated as $p_{z+}$ in the legend) and $p_z<0$ (green lines, indicated as $p_{z-}$ in the legend), and further distinguishes between spectra integrated in the whole box (solid lines) and only extracted from the reconnection downstream ($\mathcal{M}>\mathcal{M}_0$, dashed lines). We find that high-energy positrons with $p_z<0$ are mostly located within the downstream region (compare green solid and dashed lines), i.e., non-thermal positrons with $p_z<0$ are trapped in plasmoids, in analogy to 2D results \citep[see][]{petropoulou_18, hakobyan_20}. 

In contrast, a significant fraction of high-energy positrons with $p_z > 0$ reside outside the reconnection region (compare blue solid and dashed lines), and we shall call them ``free''. The fraction of free positrons is an increasing function of momentum, and for $p_z\gtrsim 100$ they are more numerous than the ones located in the reconnection downstream. The $p_{z+}$ spectrum of free positrons (dotted blue line) can be modeled as a hard power law, ${\rm d}N_{\rm free}/{\rm d}p_z \propto p_z^{-1.5}$. In Appendix~\ref{slope}, we provide an analytical justification of the measured spectral slope. The cutoff in the spectrum of $p_z>0$ positrons is much higher than for $p_z<0$ positrons, suggesting that free positrons can be accelerated to much larger energies than trapped ones, as we indeed demonstrate below.\footnote{The electron spectrum shows the opposite asymmetry: electrons with $p_z>0$ mostly reside in plasmoids, and their spectrum extends to lower momenta than for free electrons with $p_z<0$.}

The asymmetry between positrons with $p_z>0$ vs  $p_z<0$ is a unique feature of our 3D setup. In a corresponding 2D simulation (see Appendix~\ref{appd}), $p_{z+}$ and $p_{z-}$ spectra are nearly identical, and nearly all  high-energy particles reside within the reconnection downstream, as already shown by Fig.~\ref{fig:prtl_hist} (right panel).

 In the inset of Fig.~\ref{fig:spectra}, we present the box-integrated positron spectra of kinetic energy (grey) and momentum in different directions, as indicated in the legend. In contrast to the $p_z$ spectrum, there is no broken symmetry between positive and negative directions in the $p_x$ and $p_y$ spectra.
 The inset shows that the peak of the energy spectrum (grey), at $\gamma-1\sim 3$, is dominated by motions along the $x$ direction of the reconnection outflows (compare with the $p_x$ spectrum, red line). In contrast, the high-energy cutoff of the positron energy spectrum at $\gamma\sim 500$ is dominated by the  $p_{z+}$ spectrum (blue). So, the most energetic positrons move mostly along the $+z$ direction (conversely, the highest energy electrons along $-z$). We also remark that the $p_y$ spectrum (orange) reaches rather high momenta (albeit, not as high as the $p_{z+}$ spectrum). This is consistent with the trajectories of high-energy positrons that we illustrate in Sec.~\ref{3.2}.

\begin{figure*}
    \centering
    \subfloat{\stackunder{\includegraphics[width=0.48\linewidth]{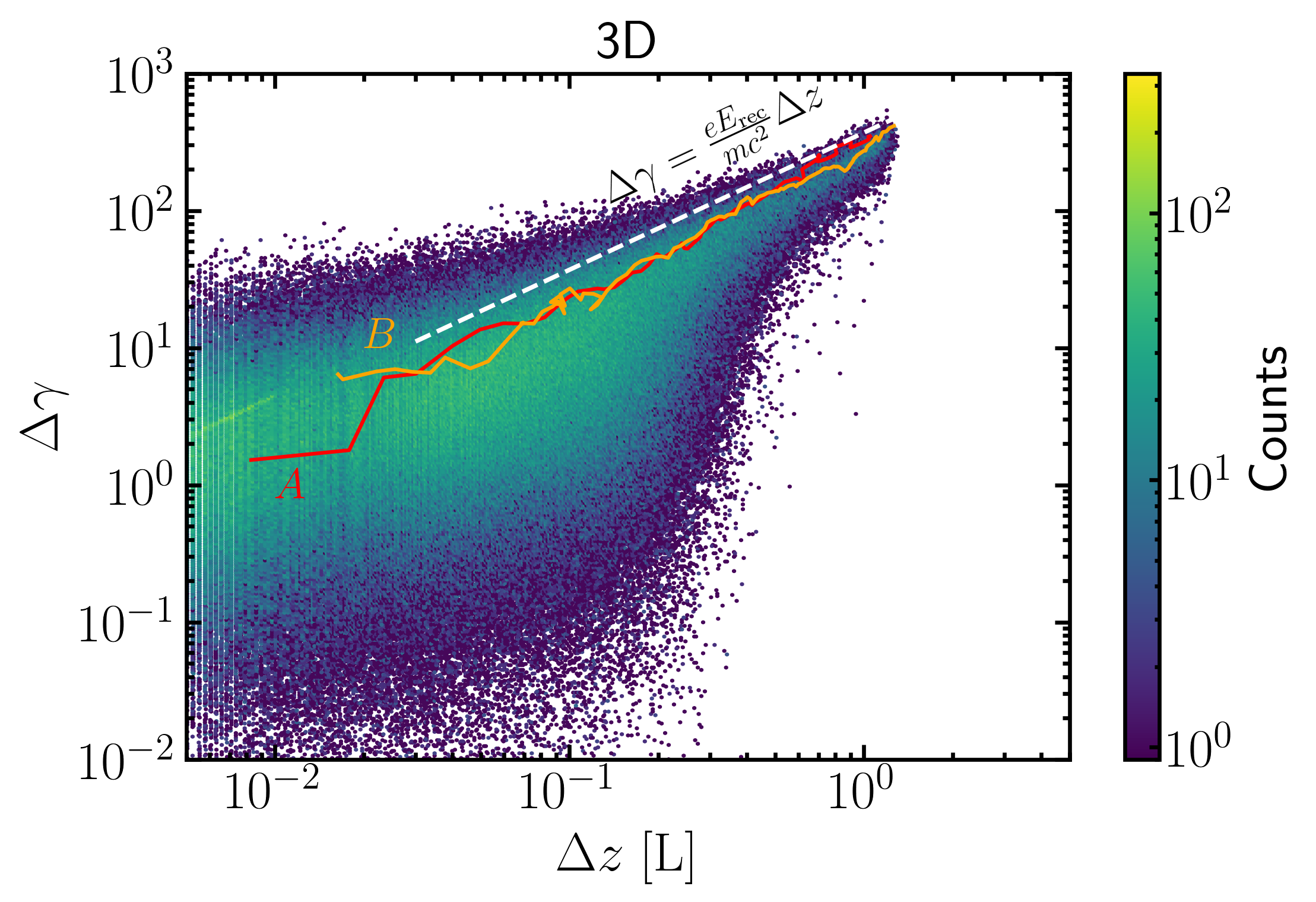}}{}}
    \qquad
    \subfloat{\stackunder{\includegraphics[width=0.48\linewidth]{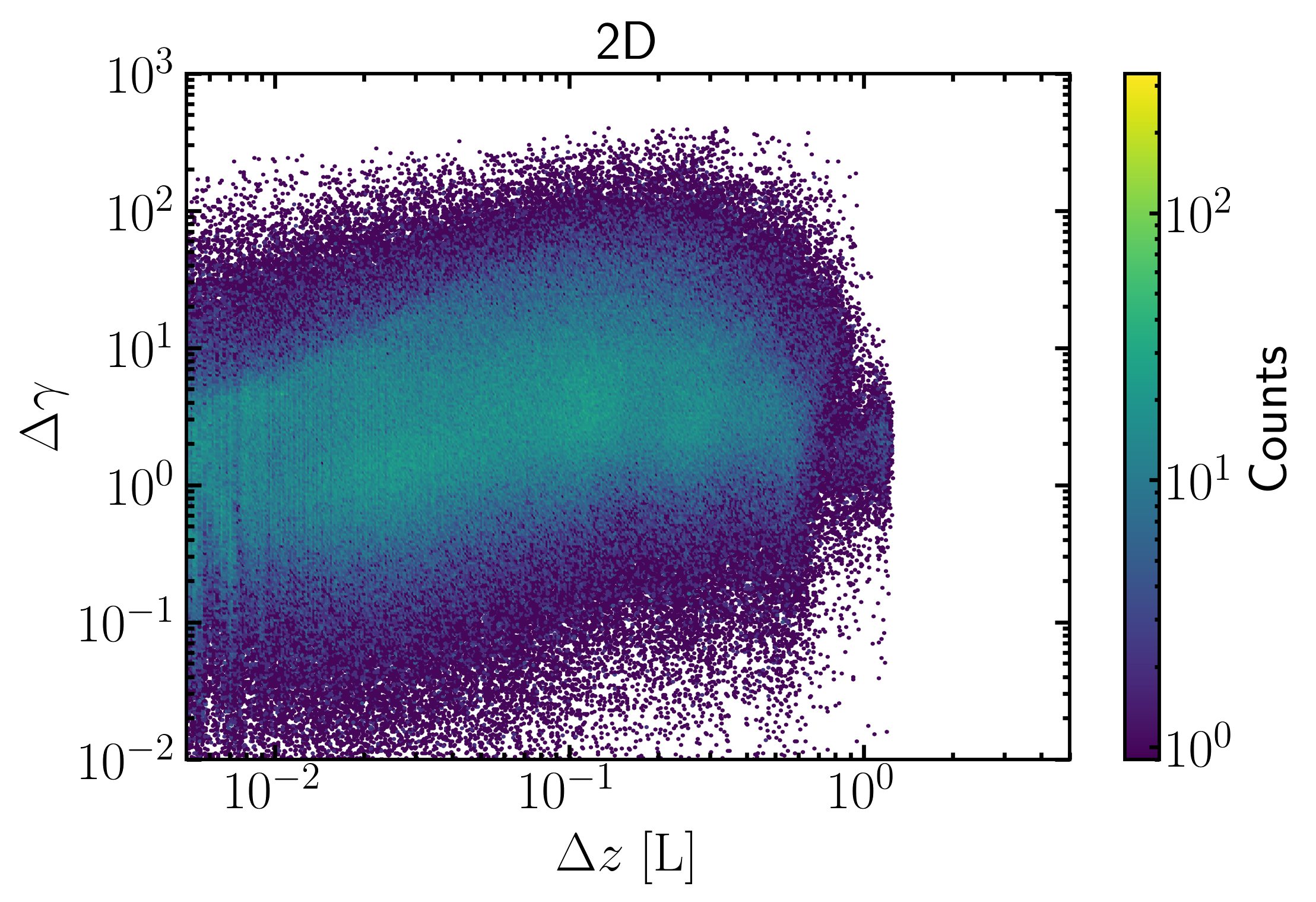}}{}}
    \caption{2D histograms of the gain in positron Lorentz factor ($\Delta \gamma$) and displacement along the $z$-axis ($\Delta z$), in 3D (left plot) and 2D (right plot). The positrons are selected at the end of the simulations ($ct/L=3.48$) and traced back to the first time they are saved. For the 2D case, the particle displacement along the $z$-axis is calculated by time integration of the $z$ velocity. 
    The relation expected from  Eq.~\ref{eq:e_est} is marked with a dashed white line in the left panel. The red and yellow lines in the left panel represent the tracks in the $\Delta \gamma-\Delta z$ plane of the two high-energy positrons shown in Fig.~\ref{fig:prtl_traj}; in this case, the differences $\Delta \gamma$ and $\Delta z$ are computed at each time with respect to the initial time when the particle Lorentz factor first exceeded $\gamma=3$.
    }
    \label{fig:dgamma_dz}
\end{figure*}

In summary, the momentum spectra in \figg{spectra} show that
most of the highest energy positrons are located in the reconnection upstream, and their momentum is dominated by the $z$ component, which is aligned with the large-scale motional electric field $\vec{E}_{\rm rec}=E_{\rm rec}\hat{z}=\eta_{\rm rec}B_0 \hat{z}$ carried by the upstream converging flows. If $\vec{E}_{\rm rec}$ is the primary agent of acceleration, we expect a linear relation between the gain in Lorentz factor ($\Delta \gamma$) and the displacement along the $z$-axis ($\Delta z$), of the form
\begin{equation}\label{eq:e_est}
    \frac{\Delta\gamma}{\Delta z} \approx \frac{e E_{\rm rec}}{mc^2} = \frac{\eta_{\rm rec}\sqrt{\sigma}\omega_{\rm p}}{c}.
\end{equation}
In Fig.~\ref{fig:dgamma_dz}, we show the relation between $\Delta \gamma$ and $\Delta z$ for a sample of $\sim 2\times 10^6$ positrons selected at the end of the simulations ($t = 3.48L/c$), for 3D (left) and 2D (right). Each particle is traced back to the first time its Lorentz factor exceeded $\gamma=3$, and its overall $\Delta \gamma$ and $\Delta z$ are computed.\footnote{In 3D, $\Delta z$ is directly recorded. In 2D, it is obtained by time integration of the $z$ velocity.} The plot only shows the quadrant with $\Delta \gamma>0$ and $\Delta z>0$, which includes most of positrons and displays the strongest difference between 2D and 3D.

For $\Delta \gamma \lesssim 100$, 2D and 3D results are similar. There appears a trend that particles gaining more energy also display a larger $z$ displacement, but the spread is quite large ($\Delta \gamma$ may vary by two orders of magnitude for the same $\Delta z$). The similarity between 2D and 3D for $\Delta \gamma \lesssim 100$ suggests that most of these particles are accelerated while trapped in plasmoids, as found in 2D simulations \citep[][]{petropoulou_18, hakobyan_20}.

The most striking difference between 2D and 3D results is in the behavior of particles experiencing large energy gains, $\Delta \gamma \gtrsim 100$. In this range ($\Delta \gamma \gtrsim 100$ and $\Delta z > 0.4\,L$), positrons from the 3D simulation follow a linear relation  $\Delta \gamma\propto \Delta z$, indicating that they are all accelerated by the same electric field. Such a branch is absent in the corresponding 2D simulation. For comparison, in the left panel of Fig.~\ref{fig:dgamma_dz} we plot with a dashed white line the expectation of \eq{e_est} for the measured  $\eta_{\rm rec} = 0.075$. The agreement of the high-energy branch in the 3D histogram with \eq{e_est} confirms that particles experiencing the largest energy gains are accelerated in the upstream by the motional electric field $E_{\rm rec}$.

We also point out the excess of positrons lying along the extrapolation of the dashed white line to low $\Delta \gamma$, in the left panel at $1\lesssim \Delta \gamma\lesssim 5$. These positrons are currently being injected into the acceleration process by the reconnection electric field, so they still obey \eq{e_est}.

\subsection{Particle Orbits}\label{3.2}

\begin{figure*}
    \centering
    \includegraphics[width=1\linewidth]{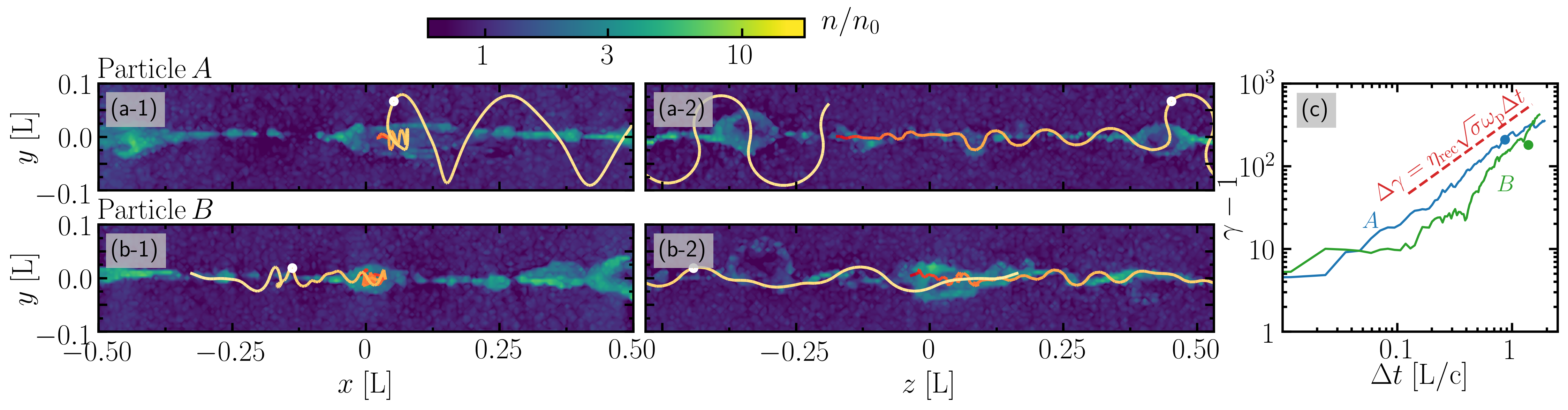}
    \caption{Trajectories of two representative positrons. For each particle, its trajectory in the $x-y$ plane is shown in the left panel, and in the $y-z$ plane in the middle panel. The color of the line represents the particle energy (from red to white as the energy increases). A white filled circle shows the position at a specific time: $t=2.28L/c$  for particle \textit{A}, corresponding to a time $\Delta t = 0.87L/c$ in the particle life; and $t=2.80L/c$ for particle \textit{B}, corresponding to $\Delta t = 1.03L/c$. The background color shows the plasma density at that same time, in the $x-y$ and $y-z$ slices where the particle is located. In the right panel, we show the particle Lorentz factor as a function of its lifetime $\Delta t$ since it first crossed a threshold $\gamma=3$. The maximum expected acceleration rate corresponding to \eq{ratemax} is shown with a red dashed line.
    }
    \label{fig:prtl_traj} 
\end{figure*}

To investigate the acceleration mechanism of the highest energy particles, we have studied the trajectory of a large number of high-energy ($\gamma>200$) positrons. We present in Fig.~\ref{fig:prtl_traj} two representative orbits.\footnote{Movies showing the orbits of positron \emph{A} and \emph{B} can be found online at \url{https://youtu.be/pjpYzw2VKe0} and \url{https://youtu.be/kOycphI0WUw}, respectively.} Their $ \Delta \gamma- \Delta z$ tracks are shown in the left panel of Fig.~\ref{fig:dgamma_dz} by the two colored lines, demonstrating that for $\Delta \gamma\gtrsim 10$ they follow a linear relation akin to \eq{e_est}.

Fig.~\ref{fig:prtl_traj} shows the particle orbits projected on the $x-y$ (left) and $y-z$ (middle) planes, as well as the particle Lorentz factor as a function of lifetime $\Delta t$ (right panel), measured since a particle first crosses the threshold $\gamma=3$. The acceleration rate due to the electric field $\vec{E}_{\rm rec}=E_{\rm rec}\hat{z}=\eta_{\rm rec}B_0 \hat{z}$ in the upstream flow can be written
\begin{equation}\label{eq:acc_rate}
    \dot{\gamma} =\frac{\Delta \gamma}{\Delta t} \approx \frac{ e E_{\rm rec}}{mc} \beta_z \approx \beta_z\eta_{\rm rec}\sqrt{\sigma} \omp,
\end{equation}
where $\beta_z$ is some time-averaged $z$ velocity in units of the speed of light. The highest acceleration rate will be achieved when $\beta_z\simeq 1$, leading to a maximal rate 
\begin{equation}\label{eq:ratemax}
\dot{\gamma}_{\rm max}=\eta_{\rm rec}\sqrt{\sigma}\omp,
\end{equation}
indicated by the dashed red line in Fig.~\ref{fig:prtl_traj} (right). We refer to this as $\dot{\gamma}_{\rm max}$, since it is the maximum acceleration rate that can be provided by the large scale electric field. Even stronger electric fields may transiently appear  within the reconnection region, which explains why some particles can temporarily experience an energization rate even larger than this value (e.g., positron \emph{B} between $\Delta t\simeq0.4$-$0.8L/c$).

We find that both positron \emph{A} and \emph{B} are injected into the acceleration process in the vicinity of an X-point in the midplane of the layer ($y=0$). Yet, at later times their histories diverge. 
Positron \emph{A} is energized at nearly the maximal rate $\dot{\gamma}_{\rm max}$ for most of its life (compare blue and dashed red lines in the right panel of Fig.~\ref{fig:prtl_traj}). Its orbit in the $y-z$ plane displays a series of quasi-periodic deflections between the two sides of the reconnection layer, as expected for Speiser motion \citep{speiser_65}. Yet, while Speiser orbits in reconnection with a weak guide field are expected to get focused towards the midplane $y=0$ \citep[e.g.,][]{cerutti_13a}, the trajectory of positron \emph{A} displays a $y$-extent increasing over time. This is caused by interactions with plasmoids, whose effect is  not taken into account in standard Speiser orbits. In fact, at the time corresponding to the white circle in Fig.~\ref{fig:prtl_traj}~(a-2), the positron has just been deflected towards the upstream by the interaction with the plasmoid located at $z\sim 0.45 L$. The positron Lorentz factor at this time is $\gamma\sim 200$ and its Lamor radius is $r_{\rm L} =\gamma m c^2/eB_0\simeq 0.08L$, which is larger than the plasmoid transverse width. It follows that the positron will not be captured by the plasmoid, but rather it is deflected away from the midplane, which allows positron \emph{A} to keep gaining energy at nearly the maximal rate, while executing a Speiser-like motion.

The orbit of particle \emph{B} is different, and more typical of the majority of high-energy positrons. It is trapped in a plasmoid in the interval $0.1L/c\lesssim \Delta t\lesssim0.4L/c$. During this stage, it moves back and forth in both  $x$- and  $z$-directions, while its Lorentz factor stays roughly constant at $\gamma\sim 20$. The positron succeeds in escaping the plasmoid at $\Delta t\sim 0.5 L/c$. After that, it experiences fast acceleration while being deflected in a Speiser-like fashion between the two converting upstream flows, similarly to positron \emph{A}. By studying a sample of $\gamma\sim 30$ particles temporarily trapped in a given plasmoid, we have found that the ones that manage to escape have typically larger $z$ velocities and are preferentially located in the plasmoid outskirts. This is expected, since such particles, by moving along $z$, will be able to successfully travel outside the plasmoid, and thus experience efficient acceleration by the upstream field. Clearly, this cannot happen in 2D, where the $z$ direction is invariant (i.e., plasmoids are infinitely long in $z$).

\begin{figure*}
    \centering
    \subfloat[\label{fig:mix_stat}]{%
        \includegraphics[width=0.319\textwidth]{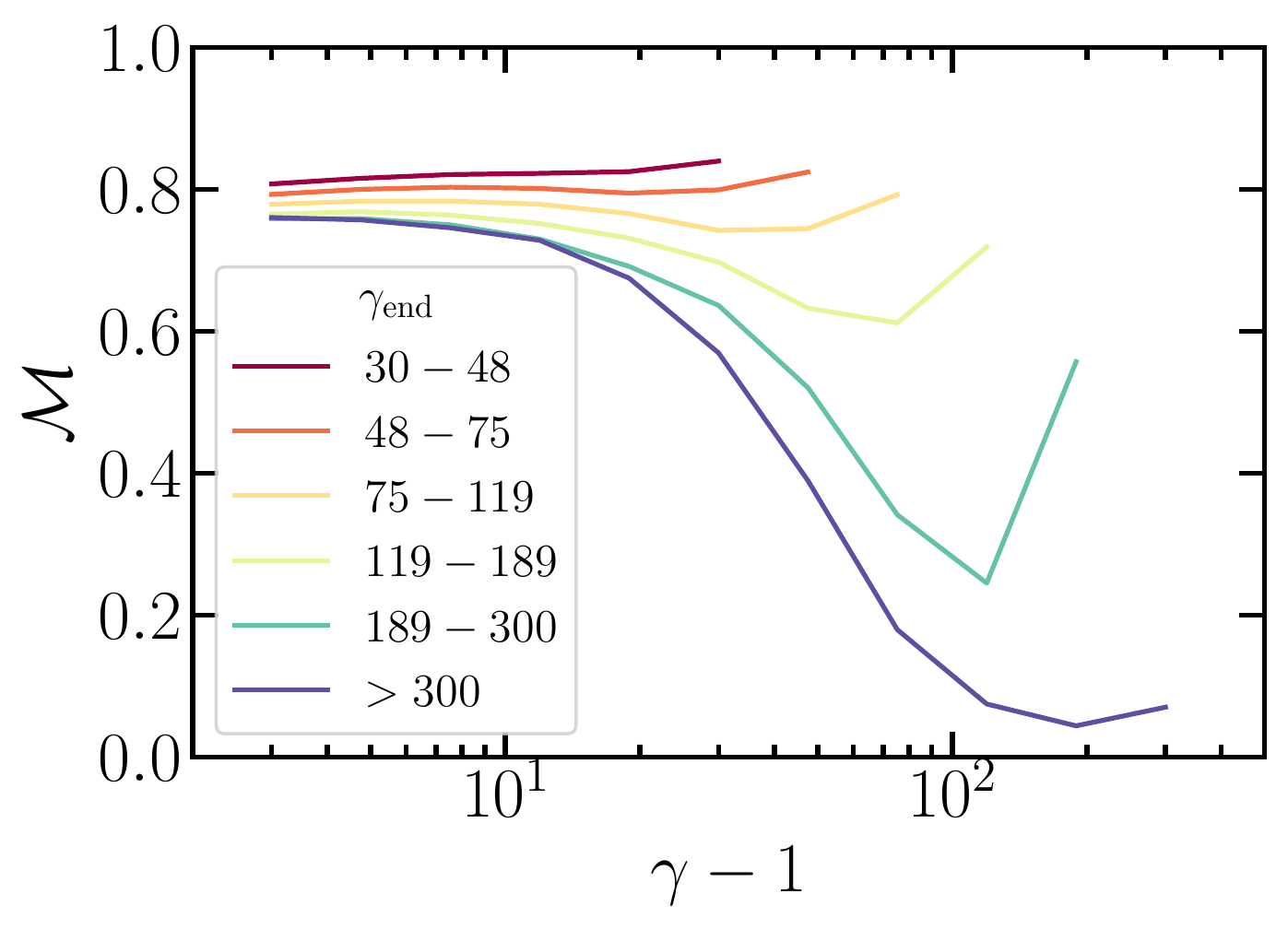}}
    \hspace{\fill}
        \subfloat[\label{fig:acc_rate_stat}]{%
        \includegraphics[width=0.322\textwidth]{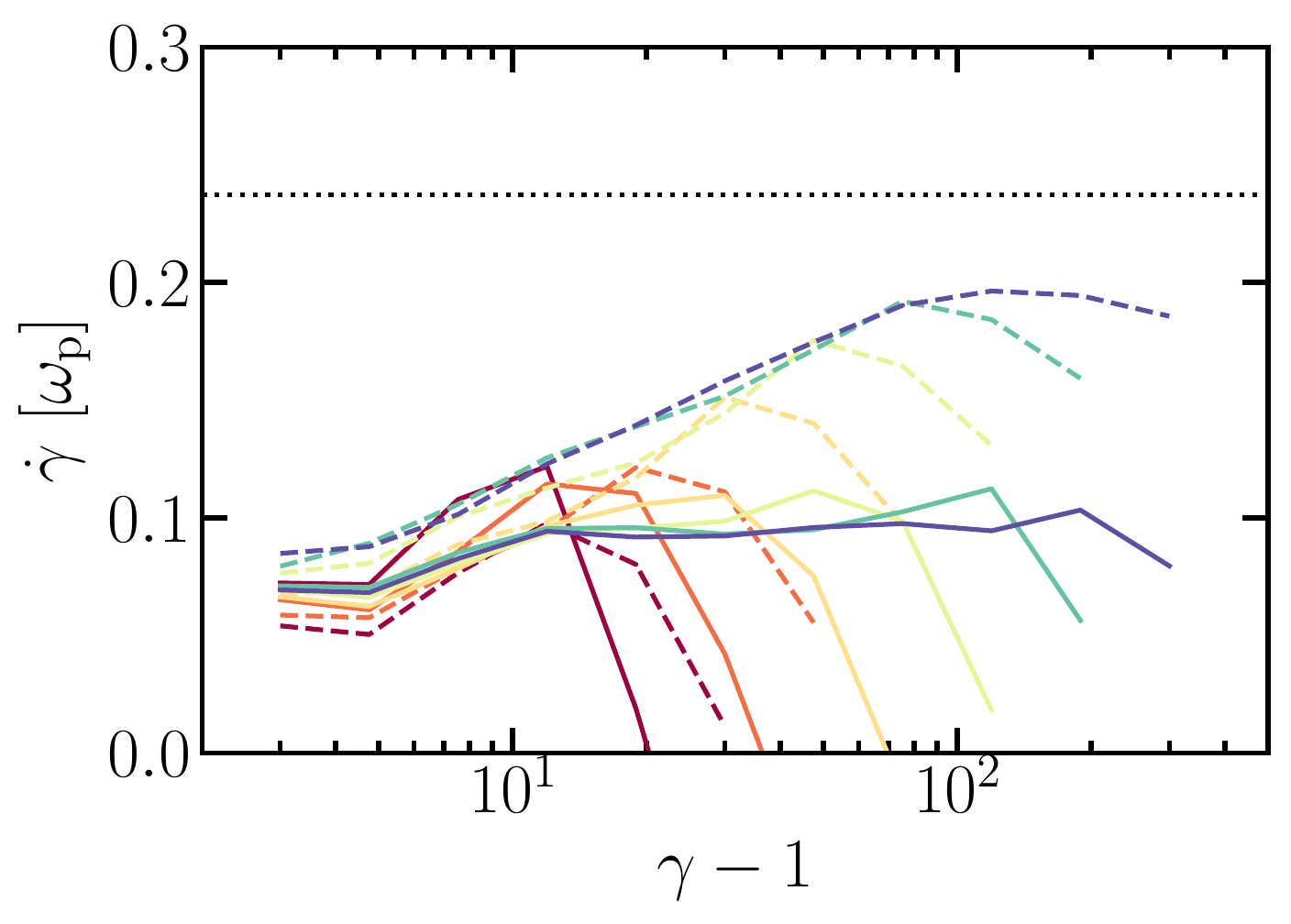}}
    \hspace{\fill}
    \subfloat[\label{fig:dgamma_frac_stat}]{%
        \includegraphics[width=0.329\textwidth]{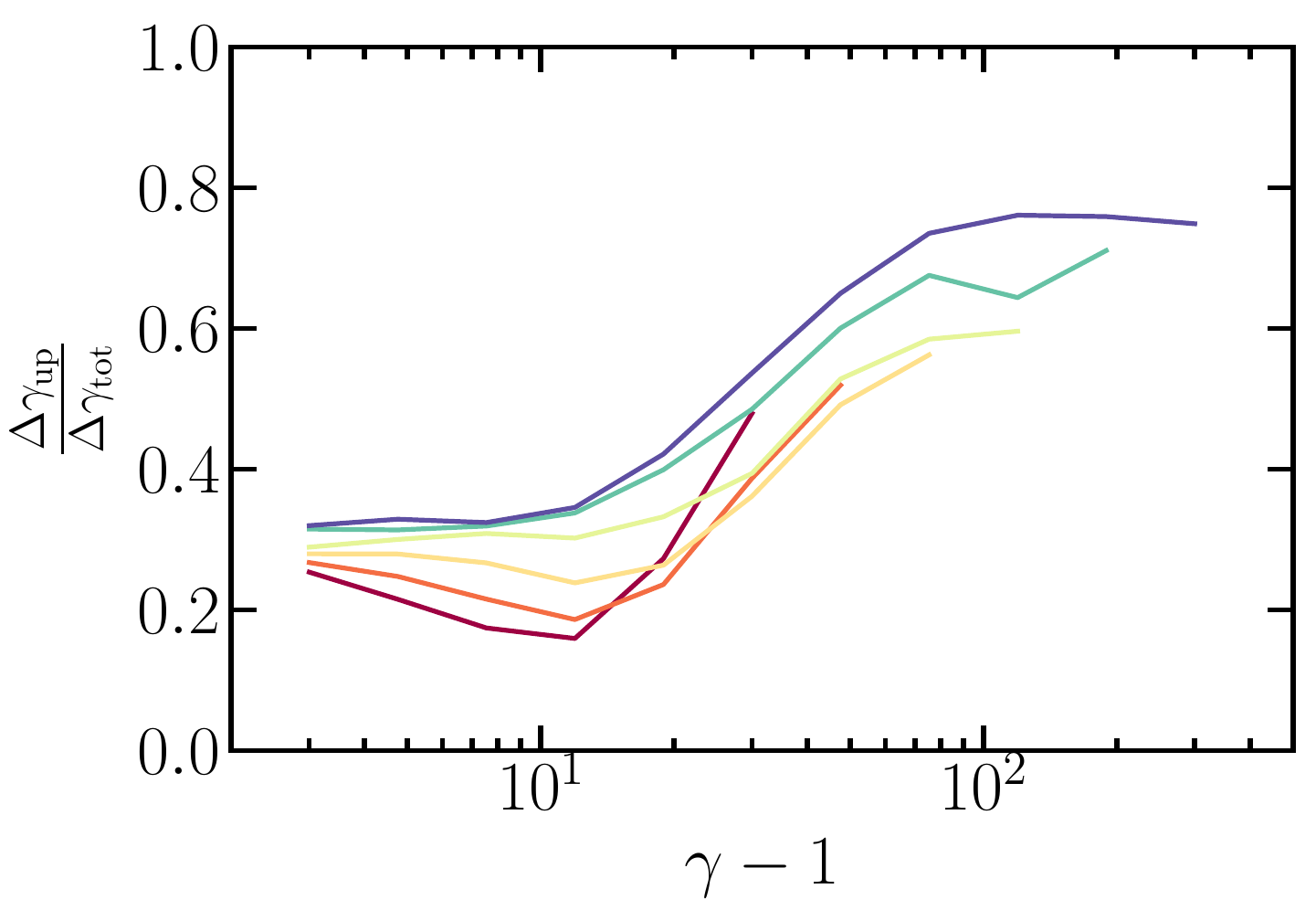}}\\
    \caption{
    Statistical assessment of the properties of accelerated positrons. We first separate the positrons in six groups, based on the largest  Lorentz factor  $\gamma_{\rm end}$ they attain in their lifetime   (see legend in the left panel). For each group, we then compute (as described in the text) the following quantities, as a function of the particle Lorentz factor $\gamma$: the median mixing factor $\mathcal{M}$ (left); the median acceleration rate $\dot{\gamma}$ (middle), distinguishing between particles in the downstream (solid) and in the upstream (dashed); the fractional energy $\Delta \gamma_{\rm up}/\Delta \gamma_{\rm tot}$ gained while in the upstream (right panel). In the middle panel we also show, as a reference, the maximum  acceleration rate quantified by Eq.~\ref{eq:acc_rate} (horizontal dotted line). 
    }
    \label{fig:stat} 
\end{figure*}

Motivated by the trajectory of particle \emph{B}, we now employ a statistical approach to further investigate the properties of accelerated particles, and in particular ascertain at which energy they are most likely to escape from plasmoids and start experiencing fast acceleration by the upstream large-scale fields. This is shown in Fig.~\ref{fig:stat}. We first separate the positrons in six groups, based on the largest  Lorentz factor they attain in their lifetime (we shall call it $\gamma_{\rm end}$, given that it is typically attained at the end of the particle life; we only consider $\gamma_{\rm end}>30$). Each group corresponds to a different color in Fig.~\ref{fig:stat}. Each of the colored curves is obtained as follows. For each particle in a given $\gamma_{\rm end}$-group, its history is followed since its birth, dividing it depending on the instantaneous  Lorentz factor (for each of the six $\gamma_{\rm end}$-groups, we employ ten $\gamma$-bins, logarithmically spaced between $\gamma=3$ and $\gamma=300$). Taking all the times when a particle lies in a given $\gamma$-bin, we compute the median mixing factor $\mathcal{M}$, median acceleration rate $\dot{\gamma}$, and the fractional energy $\Delta \gamma_{\rm up}/\Delta \gamma_{\rm tot}$ gained while in the upstream (still, while crossing the selected $\gamma$-bin). The colored lines are then computed by taking the median among particles belonging to the same $\gamma_{\rm end}$-group.

Fig.~\ref{fig:stat} (left panel) shows that at low energies ($\gamma\lesssim 20$) most of the particles reside in the downstream region, regardless of their $\gamma_{\rm end}$. In fact, the mixing fraction is $\mathcal{M}\simeq 0.8$. As particles gain energy, the median $\mathcal{M}$ of the two groups with the largest $\gamma_{\rm end}$ (green and blue lines in Fig.~\ref{fig:stat}) starts to drop, down to $\mathcal{M}\lesssim 0.1$ for the particles reaching the highest energies. As also demonstrated above, particles of high energy ($\gamma\gtrsim 100$) are preferentially located in the upstream. The transition from being trapped to breaking free appears at $\gamma\sim  3\sigma\sim 30$.


The middle panel of Fig.~\ref{fig:stat} presents the acceleration rate, distinguishing between particles in the downstream (solid lines) and in the upstream (dashed lines). The acceleration rate should be compared with the maximum rate $\dot{\gamma}_{\rm max}$ in \eq{ratemax}, which is indicated in the plot by the horizontal dotted line. We find that, regardless of $\gamma_{\rm end}$, downstream particles gain energy at a relatively slow rate, $\dot{\gamma}\lesssim 0.1\,\omp$. Particles residing in the upstream with Lorentz factors $\gamma\gtrsim 30$ --- the same threshold as derived from $\mathcal{M}$ in the left panel ---
gain energy at a faster rate, that asymptotes to  $\dot{\gamma}\sim 0.2\,\omp$ for the highest energy upstream particles.\footnote{In the highest $\gamma$-bin, all the curves bend towards slower acceleration rates. This can be simply understood as a selection bias: for a given $\gamma_{\rm end}$-group, particles in the highest $\gamma$-bin are biased towards having slower acceleration rates, otherwise they would move up in energy, and be classified in the next $\gamma_{\rm end}$-group.} This rate approaches $\simeq 0.8\,\dot{\gamma}_{\rm max}$, which implies that the highest energy particles move with an average $z$ velocity $\beta_z\simeq 0.8$ (see \eq{acc_rate}). This is in agreement with the momentum spectra presented in Sec.~\ref{3.1}, i.e., the highest energy particles preferentially move in the $z$ direction.


The right panel of Fig.~\ref{fig:stat} shows the fraction $\Delta \gamma_{\rm up}/\Delta \gamma_{\rm tot}$  of energy acquired in the upstream, while traversing a given $\gamma$-bin. Regardless of the $\gamma_{\rm end}$-group, we find that this is an increasing function of $\gamma$, reaching $\sim 80\%$ for the highest energy particles. Again, the transition to the stage when acceleration is dominated by the upstream motional field occurs at $\gamma\gtrsim 30$, the same threshold already derived from the left and middle panels. So, we conclude that most particles ending up with high energies escape from plasmoids at $\gamma\sim 3 \sigma\sim 30$, at which point their energization starts to be dominated by the large-scale upstream field.




\subsection{Dependence on the Domain Size}\label{3.3}

\begin{figure*}
    \centering
    \subfloat[\label{fig:g_max}]{%
        \includegraphics[width=0.31\textwidth]{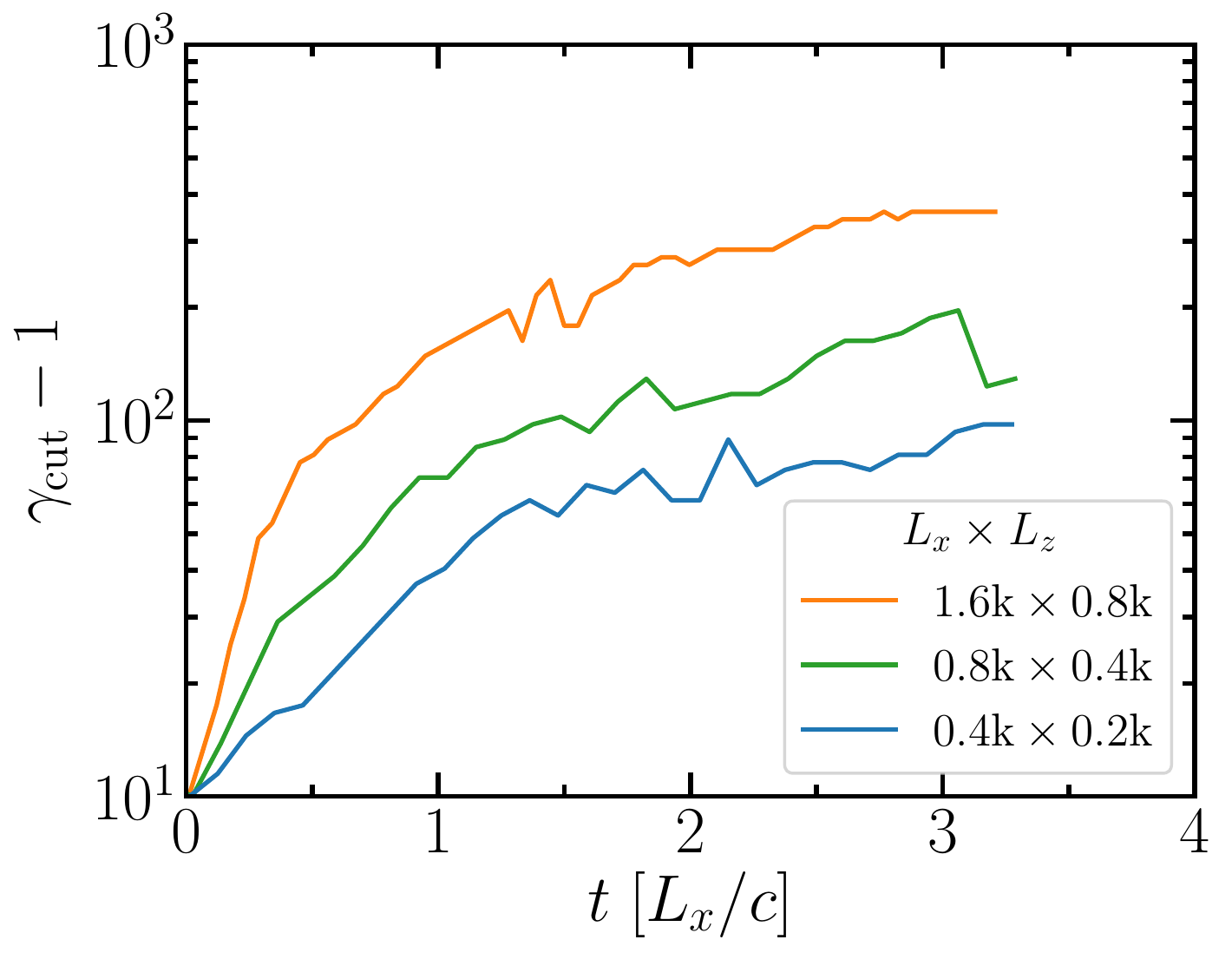}}
    \hspace{\fill}
    \subfloat[\label{fig:ratio_x} ]{%
        \includegraphics[width=0.33\textwidth]{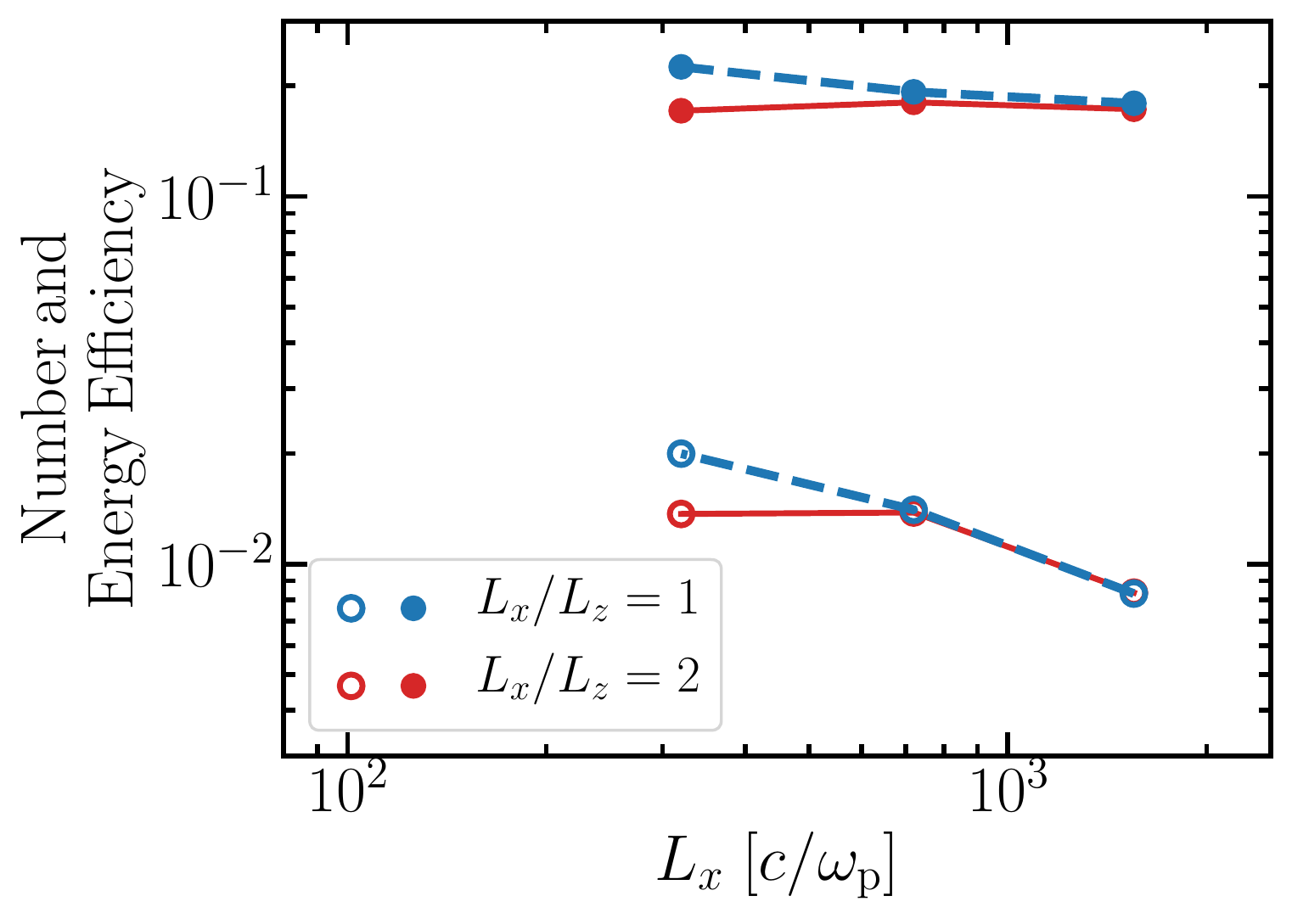}}
    \hspace{\fill}
    \subfloat[\label{fig:ratio_z}]{%
        \includegraphics[width=0.33\textwidth]{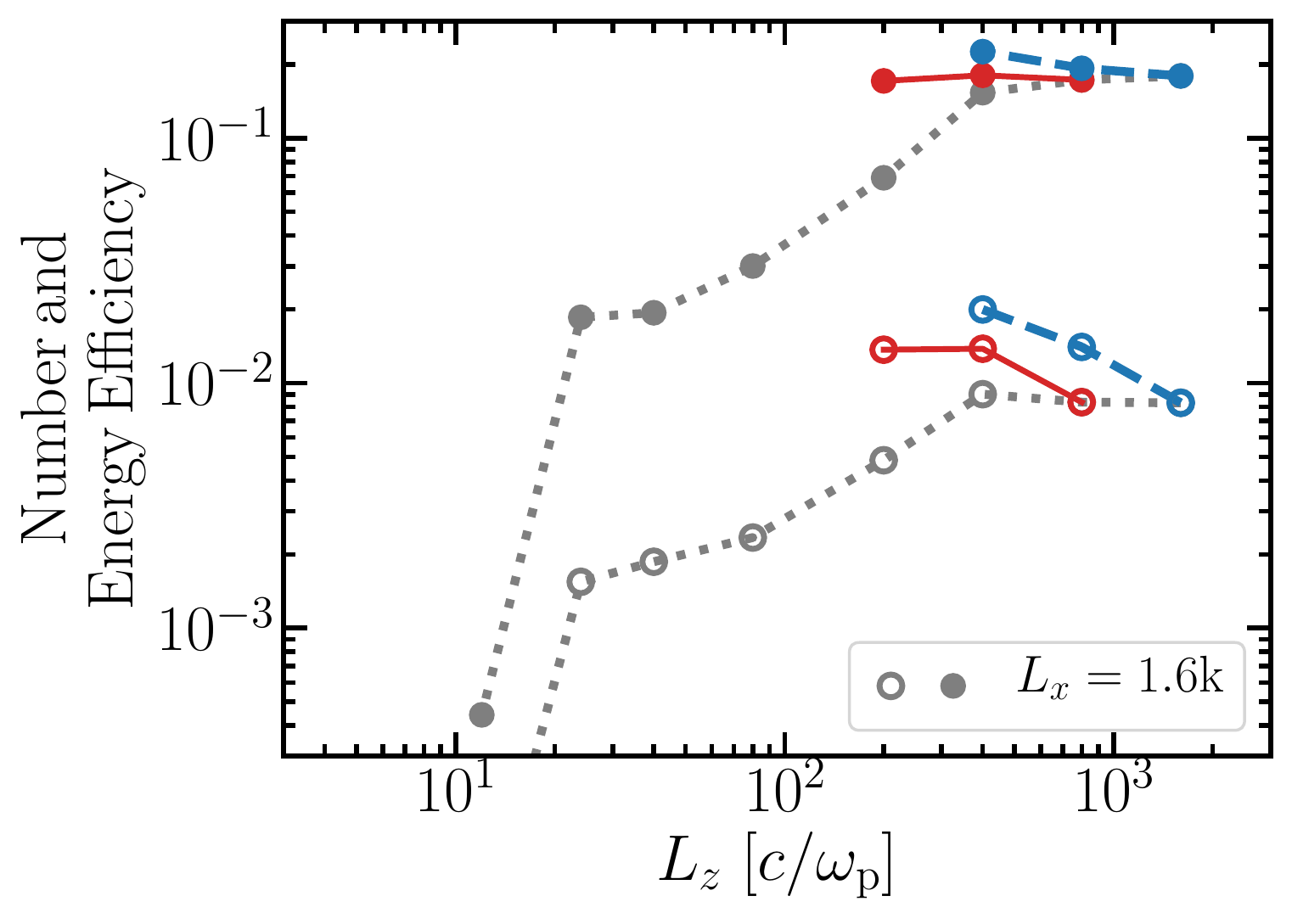}}\\
    \caption{Dependence on the box size. The unit of $L_x$ and $L_z$ is the plasma skin depth ($\comp$).
    (a) The energy cutoff $\gamma_{\rm cut}$ for free particles as a function of time for boxes with $L_x = 2\,L_z$ (see legend). (b) Number and energy efficiency for boxes with different $L_x$, at fixed aspect ratio $L_x/L_z=1$ (blue) and $L_x/L_z=2$ (red). Empty circles show the number efficiency, defined as $N_{\rm free}/N_{\rm rr}$, where $N_{\rm free}$ and $N_{\rm rr}$  are respectively the number of free particles and of particles in the reconnection region; filled circles show the energy efficiency $E_{\rm free}/E_{\rm rr}$.  
    (c) Number (empty grey circles) and energy (filled grey circles) efficiency for boxes with different $L_z$, at fixed $L_x\sim 1600\comp$. Blue and red points are the same as in (b).
    }
    \label{fig:scan_box}
\end{figure*}

\begin{table}[] 
\begin{threeparttable}
\begin{tabularx}{\columnwidth}{cccc}
\hline\hline
box size & $\Delta_{\rm init}/[\comp]$ & $N_{\rm free}/N_{\rm rr}$ & $E_{\rm free}/E_{\rm rr}$ \\ 
\hline
$1.6k\times0.8k$ & $500$ & $0.006$($0.008$)\tnote{$\dagger$} & $0.131$($0.173$) \\ 
\hline
\multirow{2}{*}{$0.8k\times0.4k$} & $500$ & $0.012$ & $0.161$ \\
& $250$ & $0.015$ & $0.200$ \\ 
\hline
\multirow{3}{*}{$0.4k\times0.2k$} & $500$ & $0.012$ & $0.149$ \\
& $250$ & $0.016$ & $0.199$ \\
& $125$ & $0.013$ & $0.166$ \\ 
\hline
$1.6k\times1.6k$ & $500$ & $0.008$ & $0.179$ \\ 
\hline
\multirow{2}{*}{$0.8k\times0.8k$} & $500$ & $0.014$ & $0.188$ \\
& $250$ & $0.014$ & $0.197$ \\ 
\hline
\multirow{3}{*}{$0.4k\times0.4k$} & $500$ & $0.022$ & $0.224$ \\
& $250$ & $0.023$ & $0.250$ \\
& $125$ & $0.015$ & $0.201$ \\
\hline
$1.6k\times0.4k$ & $500$ & $0.009$($0.012$)\tnote{$\ddagger$} & $0.153$($0.199$) \\
\hline
\end{tabularx}
\begin{tablenotes}\footnotesize
\item[$\dagger$] The results in  parentheses are from a simulation with the same parameters, but with four particles per cell.
\item[$\ddagger$] The results in  parentheses are from a simulation with the same parameters, but with guide field $B_g = 0$.
\end{tablenotes}
\caption{Number and energy efficiency for the population of high-energy free particles. The number (energy, respectively) efficiency is the ratio of the number (energy) of free particles normalized to the number (energy) of particles in the reconnection region. The unit of length for the box size (leftmost column) is the plasma skin depth ($\comp$). 
} \label{tab:box}
\end{threeparttable}
\end{table}

In this subsection, we investigate the dependence of the properties of high-energy ``free'' particles on the size of the computational domain, in order to extrapolate our conclusions to larger (fluid) scales. Free particles are defined such that they reside in the upstream, with mixing parameter $\mathcal{M}<0.3$. We also require that they have $\gamma>10$, to exclude the cold upstream particles that have yet to reach the reconnection region. Our results are presented in Fig.~\ref{fig:scan_box} and Tab.~\ref{tab:box}.


In the left panel, we show the cutoff Lorentz factor $\gamma_{\rm cut}$ of
free particles, as a function of time (horizontal axis) and box size (different colors, as indicated in the legend). The cutoff Lorentz factor is obtained by calculating the location of the peak of $(\gamma-1)^3{\rm d}N_{\rm free}/{\rm d}\gamma$. As described in \citet{petropoulou_18}, this is generally a good proxy for the location of the exponential cutoff of the spectrum. We find that $\gamma_{\rm cut}\propto L_x$. The proportionality constant is such that the Larmor radius of particles with Lorentz factor $\gamma_{\rm cut}$ is $r_{\rm L}(\gamma_{\rm cut})\sim 0.2 \,L_{x}$, regardless of $L_x$. This is expected, since particles accelerated near the maximal rate in \eq{ratemax} over the typical advection time $\sim L_x/c$ will obtain a Larmor radius $r_{\rm L}\sim \eta_{\rm rec}L_x\sim 0.1 L_x$.
This can be phrased as a ``Hillas criterion'' for relativistic magnetic reconnection.

We also calculate the energy efficiency $E_{\rm free}/E_{\rm rr}$ (filled circles), where $E_{\rm free}$ and $E_{\rm rr}$ are respectively the energy content of free particles and of particles in the reconnection region. The number fraction $N_{\rm free}/N_{\rm rr}$ (open circles) is obtained in a similar way. We examine their dependence on the $x$ length of the box in the middle panel (at fixed aspect ratio $L_x/L_z$, blue for $L_x/L_z=1$ and red for $L_x/L_z=2$) and on the $z$ length in the right panel, for fixed $L_x\sim 1600\comp$ (grey points).
As shown in Fig.~\ref{fig:ratio_x}, $E_{\rm free}/E_{\rm rr}$ is nearly independent of $L_x$. This demonstrates that, regardless of the box size, free particles carry a constant fraction ($\sim 20\%$) of the post-reconnection particle energy. Given that $\gamma_{\rm cut}\propto L_x$ and that the spectrum of free particles is hard, $dN_{\rm free}/d\gamma\propto \gamma^{-1.5}$, this implies that their number fraction needs to decrease with increasing box size, as indeed confirmed by Fig.~\ref{fig:ratio_x} (open circles).

Fig.~\ref{fig:ratio_z} shows that convergent 3D results are obtained only if the box is sufficiently extended in the $z$ direction. For our reference case with $L_x\sim 1600\comp$, convergent 3D results are obtained for $L_z\gtrsim 400\,\comp\sim L_x/4$. This may be due to the requirement that the $z$ extent of the largest plasmoids, $\sim 0.1 L_x$ (assuming spherical plasmoids), be smaller than the box length along $z$, i.e.,  $z$ invariance should be broken even for the largest plasmoids. Fig.~\ref{fig:ratio_z} also shows that the 2D limit is approached for $L_z\lesssim 20\,\comp$, such that even small plasmoids do not fit within the vertical extent of the box. 

\section{Summary and Discussion}\label{4}
In this work, we performed large-scale 3D PIC simulations of relativistic reconnection in a $\sigma=10$ electron-positron plasma.
We found that a fraction of particles with $\gamma\gtrsim 3\sigma$ can ``break free'' from plasmoids by moving along $z$ and then experience the large-scale motional electric field in the upstream region. This process cannot be captured by 2D simulations, which are invariant along the $z$ direction.
The free particles preferentially move along $z$ and are accelerated linearly in time ($\gamma\propto t$) while undergoing Speiser-like deflections by the converging upstream flows, as already hypothesized by \citet{giannios_10}.
Their spectrum is  hard and can be modeled as a power law $dN_{\rm free}/d\gamma\propto \gamma^{-1.5}$ --- in Appendix \ref{slope}, we analytically justify the value of the power-law slope. The free particles account for $\sim 20\%$ of the dissipated magnetic energy, independently of domain size.

{The acceleration rate of the particles, in this mechanism,
is closely connected to the reconnection speed. 
Its accurate description, therefore, relies on 
the reconnection system having reached a statistical steady
state. To this end, our adopted boundary conditions    
are of crucial importance. By adopting continuous
injection of plasma (and magnetic flux) in the far upstream and outflow boundaries in the 
reconnection exhaust direction, the system can be followed for many Alfv\'en crossing times
after it has reached a statistical steady state. We find that the most energetic particles take
a few Alfv\'en crossing times to reach their maximum energy. In contrast, the more commonly adopted triple periodic
boundaries can only study reconnection transiently  and may not be able to capture
this mechanism accurately. In addition, periodic boundaries do not allow plasmoids to escape, so the largest plasmoids can grow up to a size comparable to the system length. This would artificially enhance the rate at which high-energy free particles get captured back by plasmoids.}



We find that the particle acceleration rate is
\begin{equation}
    \dot{\gamma} \sim \frac{\eta_{\rm rec} eB}{mc} \beta_z,
\end{equation}
where $B$ is the magnetic field in the upstream and $\eta_{\rm rec}\beta_z\sim 0.06$ as determined by our simulations. As far as we can infer from our simulations, the maximum 
energy achievable by this process is determined by either radiative losses or by the size of the reconnection region. If synchrotron cooling
is the dominant energy loss, then the particles are accelerated until the energy gain rate $\dot{\gamma} m c^2$ is balanced by the synchrotron loss rate
$(4/9)e^4B^2\gamma^2/m^2c^3$, resulting in a maximum $\gamma_{\rm max}$:
\begin{equation}\label{eq:gmax_red}
\gamma_{\rm max}=\sqrt{\frac{9\beta_z\eta_{\rm rec}m^2c^4}{4e^3B}}.
\end{equation}
The corresponding synchrotron emission energy is $E_{\rm syn}=\gamma_{\rm max}^2Be\hbar/mc=160\beta_z\eta_{\rm rec}$~MeV$\sim 10$~MeV, for electrons, i.e., about one order of magnitude below the well-known burnoff limit \citep{dejager_harding_92}. We remark that in this argument we have assumed that the accelerated particles have large pitch angles (i.e., the angle between the particle velocity and the magnetic field), as indeed observed in our simulations. Yet, if this acceleration mechanism were to operate also in the limit of strong guide fields, the accelerated particles would likely have small pitch angles, which would comparatively reduce their synchrotron  losses.\footnote{On the other hand, a strong guide field would also help in enforcing invariance along the $z$ direction (i.e., plasmoids are likely to be very elongated in $z$), so our proposed mechanism may turn out to be less efficient.} 

If radiative losses are negligible, the maximum energy is only limited by the size of the reconnecting system.
For a given length of the reconnection layer $L_z$ in the $z$ direction, the maximum Lorentz factor a particle can reach can be estimated as: 
\begin{equation}\label{eq:gmax}
    \gamma_{\rm max} = L_{\rm z}e\eta_{\rm rec}B/mc^2.
\end{equation}
The particle motion along the $x$-direction of reconnection outflows is unlikely to constrain the maximum energy, since we have shown that the accelerated particles mostly move along the $z$ direction, so their escape time along $x$ is likely longer than along $z$.

Although in this work we have only focused on electron-positron plasmas, our results may still be applicable to electron-proton cases, since high-$\sigma$ reconnection is virtually identical in pair plasmas and electron-proton plasmas. 
Protons are much less affected by cooling as compared to leptons and may escape from plasmoids with higher efficiency. Therefore, given that we have neglected cooling losses, our results may be most applicable to protons in astrophysical systems where radiative lepton losses are severe. Nevertheless, in less extreme environments (e.g., the emission zone of a blazar jet), even leptons should be able to able to escape small plasmoids and participate in this acceleration process. Therefore, one may expect that both leptonic and hadronic signatures will be affected by the acceleration mechanism we have discussed here. 


The sources and acceleration mechanism of UHECRs with energies between $\sim 10^{18}$ and $\sim10^{20}$~eV are still under debate. Relativistic jets launched by GRBs \citep{milgrom_1995, waxman_1995} and AGNs \citep{halzen_02} have been proposed as sources of UHECRs. Magnetic reconnection  taking place in the magnetically dominated plasma of these jets may be a promising accelerator of UHECRs \citep{giannios_10}.

The rest-frame magnetic field strength can be estimated by the Poynting luminosity of the jet $\mathscr{L}_{\rm P}$, the bulk Lorentz factor $\Gamma$, and the distance $R$ from the central engine (see, e.g., \citet{giannios_10}):
\begin{equation}
B\simeq \frac{\mathscr{L}_{\rm P}^{1/2}}{c^{1/2}R\Gamma}.
\end{equation}
Combining with Eq.~\ref{eq:gmax}, we can estimate the maximum energy that a proton can be accelerated to:
\begin{equation}
	E_{\rm max} = \eta_{\rm rec}  \mathscr{L}_{\rm P}^{1/2} c^{-1/2} R^{-1} L e,
\end{equation}
where $L$ is the length scale of the reconnection region. We also introduce a factor of $\Gamma$ when we boost from the jet frame to the observer frame.

For long-duration GRBs, the (isotropic equivalent) energy they release in gamma rays is around $10^{53}$~erg in a duration of about $10$~s. This gives a lower limit for their Poynting luminosity  of $\mathscr{L}_{\rm P}\sim 10^{52}~{\rm erg\,s^{-1}}$, since the energy conversion efficiency to gamma rays needs to be below unity. The size of the reconnection region can be estimated as $L\sim R/\Gamma$, and $\Gamma$ usually varies from 100 to 1000 in GRB jets. Therefore, the maximum energy that a particle can be accelerated to is $E_{\rm max} \approx 2\times 10 ^{20} \eta_{\rm rec,-1}\mathscr{L}_{\rm P, 52}^{1/2} \Gamma_2^{-1} {\;\rm eV}$, where $\eta_{\rm rec,-1} = \eta_{\rm rec}/0.1$, $\mathscr{L}_{\rm P, 52} = \mathscr{L}_{\rm P}/10^{52}{\rm erg \,s^{-1}}$, and $\Gamma_2 = \Gamma/10^2$.

A powerful AGN jet can reach a luminosity of $10^{48}~{\rm erg\,s^{-1}}$. They usually have bulk Lorentz factors of  $\Gamma \sim 3-30$. Using these values, we estimate that protons accelerated by reconnection in AGN jets can reach energies $E_{\rm max}\sim 6\times10^{19} \eta_{\rm rec,-1}\mathscr{L}_{\rm P, 48}^{1/2} \Gamma_{0.5}^{-1} {\;\rm eV}$ where $\mathscr{L}_{\rm P, 48} = \mathscr{L}_{\rm P}/10^{48}{\rm erg\,s^{-1}}$ and $\Gamma_{0.5} = \Gamma/\sqrt{10}$.

Both jets are then capable of accelerating protons (or heavier nuclei in AGN jets) to $10^{18}$~eV and even to the highest energies that have been observed so far, $10^{20}$~eV. Though we do not consider constrains imposed by cooling losses (see, e.g., \citet{giannios_10} for further discussion), our analysis demonstrates that relativistic reconnection is, in principle, a promising way to produce UHECRs.


\acknowledgements
HZ and DG acknowledge support from the NASA ATP NNX17AG21G, the NSF AST-1910451, and the NSF AST-1816136 grants. LS acknowledges support from the Sloan Fellowship, the Cottrell Scholars Award, NASA ATP NNX17AG21G and NSF PHY-1903412. The simulations have been performed at Columbia (Habanero and Terremoto), with NERSC (Cori) and NASA (Pleiades) resources.


\appendix

\section{2D Particle Spectra}\label{appd}

\begin{figure}
    \includegraphics[width=1\columnwidth]{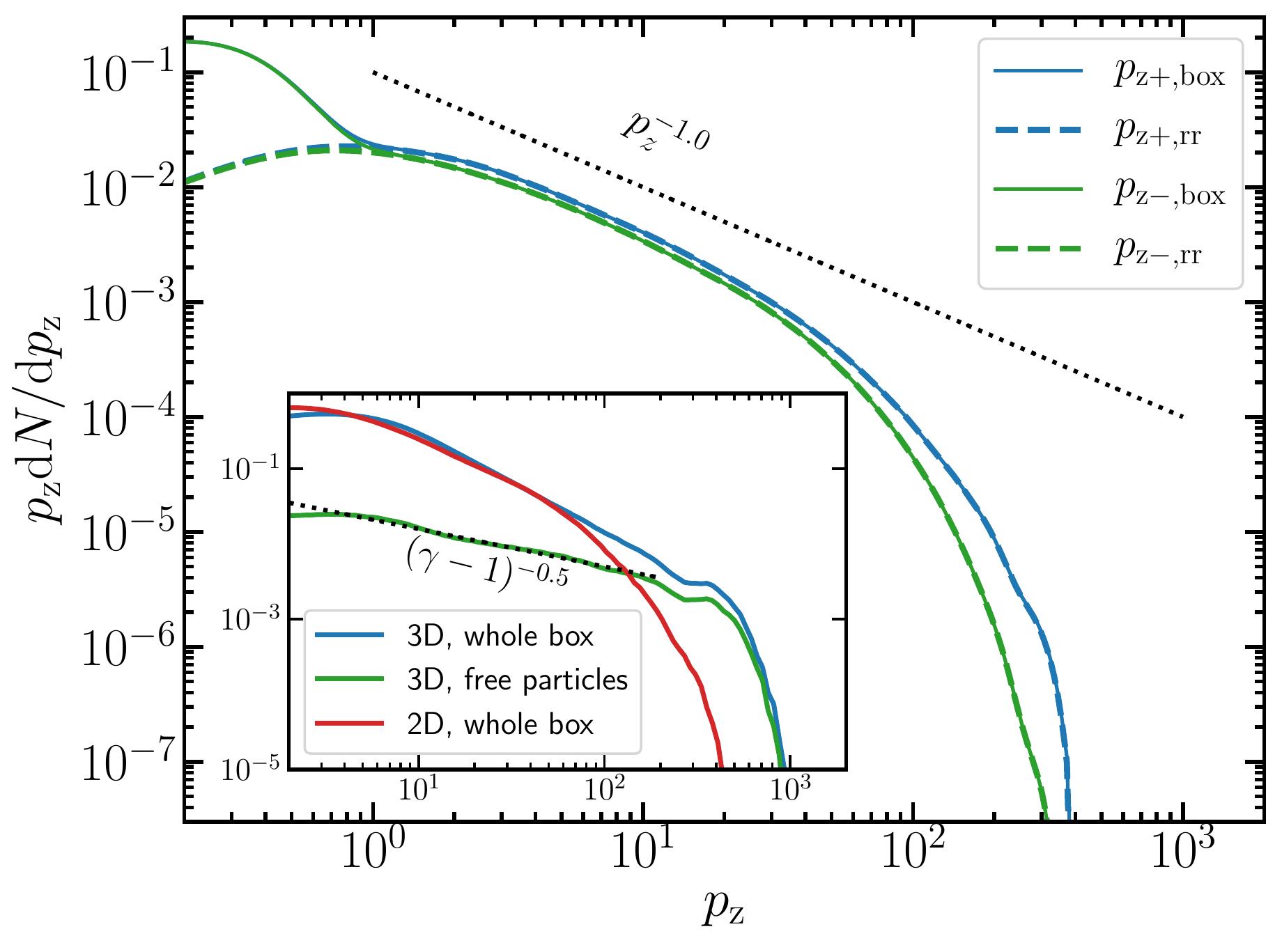}
    \caption{The main panel is as in Fig.~\ref{fig:spectra}, but for a 2D simulation. We display the positron time-averaged (between $t = 3.34L/c$ and $3.56L/c$) spectra of the momentum along $+z$ (blue lines) and $-z$  (green lines). 
    Solid lines refer to the whole box, while dashed lines to the particles in the downstream.  In the inset, we show the particle energy spectrum $(\gamma-1){\rm d}N/{\rm d}\gamma$ from the whole simulation box in the 2D simulation (red line) and in the 3D simulation (blue), as well as the spectrum of free particles from the 3D simulation (green). The latter can be fit as a power law ${\rm d}N/{\rm d}\gamma\propto (\gamma-1)^{-1.5}$, as indicated by the dotted black line in the inset. 
    }
    \label{fig:spect_2d}
\end{figure}

In Fig.~\ref{fig:spectra}, we have shown the energy and momentum spectra extracted from the 3D simulation. For comparison, in Fig.~\ref{fig:spect_2d} we show the $z$-momentum spectrum from the corresponding 2D run. In agreement with earlier 2D studies, the particle spectrum is non-thermal. Unlike the 3D case, where the cutoff in the $p_{z+}$ spectrum is much larger than in the $p_{z-}$ spectrum, here the two are roughly comparable, though particles moving along $+z$ extend to slightly larger energies, and dominate in number at high energies. In addition, as discussed in the main text, at $\gamma>2$ the spectra from the whole box are coincident with corresponding spectra extracted from the reconnection region alone.

In the inset, we compare energy spectra between 2D and 3D simulations. The 3D spectrum has a higher cutoff than the 2D one and it is dominated at high energies by particles outside the reconnection region. It indicates once again that  acceleration outside of plasmoids plays an important role for the highest energy particles. 


\section{The Power-Law Slope of Free Particles}\label{slope}

\begin{figure}
    \includegraphics[width=1\columnwidth]{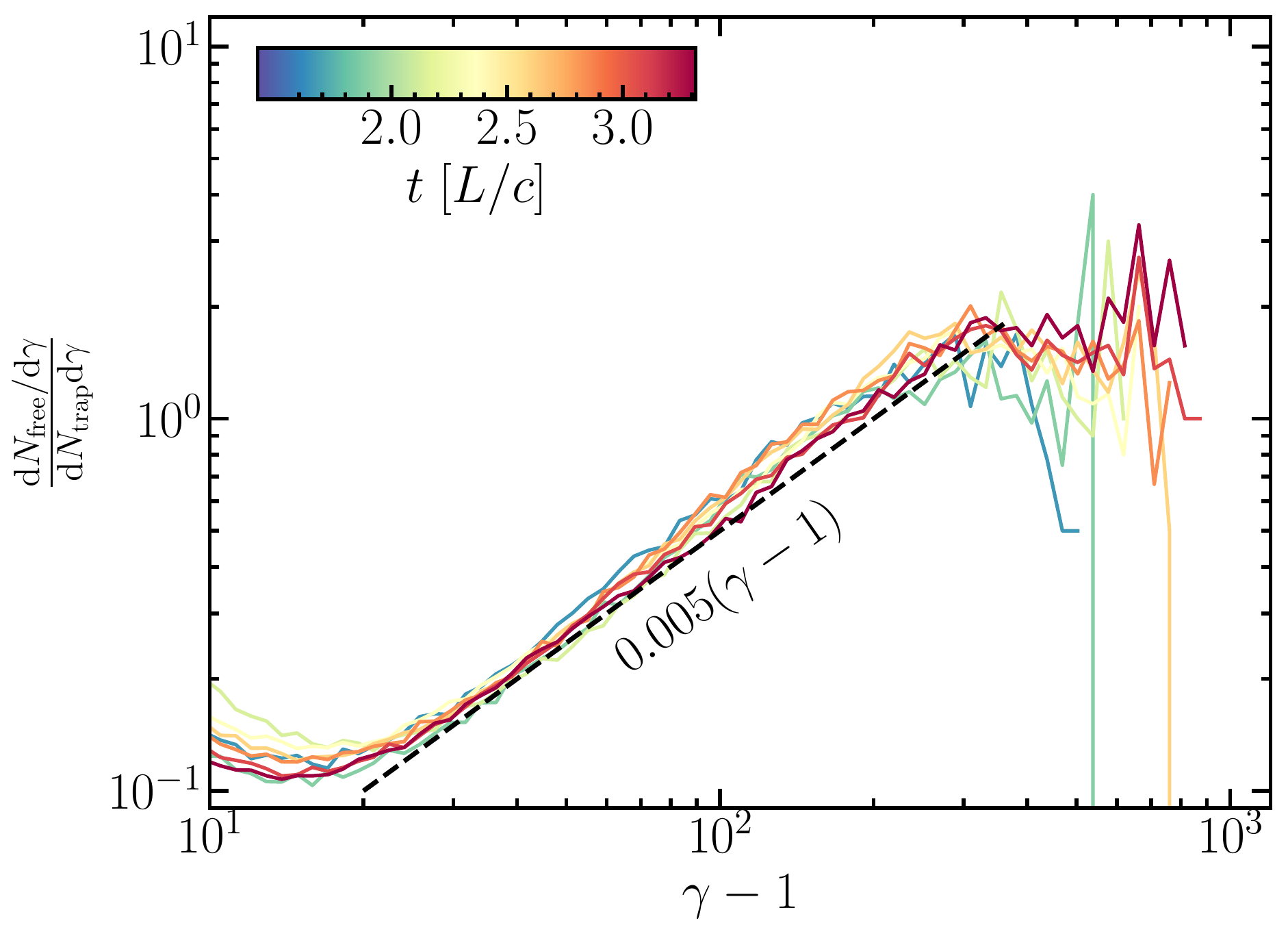}
    \caption{The ratio of the number of free particles to the number of trapped particles. The trapped particles at a given time are defined as the ones that are currently trapped, but were free at some point in the previous $\sim 0.25 L_x/c$.
    }
    \label{fig:ratio_n_free_n_trap}
\end{figure}

In the main body of the paper, we have demonstrated that the spectrum of free particles can be modeled as a power law $f={\rm d} N_{\rm free}/{\rm d} \gamma\propto \gamma^{-1.5}$ followed by a cutoff, which scales linearly with the system size. In this Appendix, we aim at providing a theoretical framework to interpret the value of the power-law slope.\footnote{We remark that, as shown in our paper, the acceleration mechanism of free particles is distinct from the one of particles accelerated in the reconnection layer, whose spectral shape has been discussed by, e.g., \citet{guo_14,uzdensky_20}.} In steady state, the distribution of free particles will follow 
\begin{equation}\label{eq:fp}
    \frac{\partial}{\partial \gamma} \left(\frac{\gamma}{t_{\rm acc}}f\right) + \frac{f}{t_{\rm esc}} = Q_0 \delta(\gamma-3 \sigma),
\end{equation}
where both the acceleration time $t_{\rm acc}$ and the escape time $t_{\rm esc}$ generally depend on the particle Lorentz factor. In the equation above, $Q_0$ quantifies the particle injection rate, which we have assumed to happen at a fixed energy of $\gamma=3\sigma$, in agreement with the results obtained in the main body of the paper.

For free particles, the acceleration time can be calculated as $t_{\rm acc} \equiv \gamma/\dot{\gamma} = \gamma/(\eta_{\rm rec}\beta_z\sqrt{\sigma}\omp)$, where our results yield $\eta_{\rm rec}\beta_z\simeq 0.06$. Free particles are accelerated while residing in the upstream. When they get captured and trapped by plasmoids, they effectively escape the accelerator, so the escape time $t_{\rm esc}$ from the acceleration region is, for free particles, the time spent in the upstream before getting trapped.\footnote{For simplicity, we assume that no significant energization occurs while trapped in plasmoids.}

In order to compute the escape time of free particles, we assume a steady state scenario: the rate at which free particles are trapped in plasmoids balances the rate at which trapped particles advect out of the domain, so 
\begin{equation}\label{eq:eq1}
t_{\rm esc}=t_{\rm adv}\frac{dN_{\rm free}/d\gamma}{dN_{\rm trap}/d\gamma} 
\end{equation}
Fig.~\ref{fig:ratio_n_free_n_trap} shows, at different times, the ratio of the number of free particles to the number of trapped particles. The trapped particles at a given time are defined as the ones that are currently trapped, but were free at some point in the previous $\sim 0.25 L_x/c$ (our results are not appreciably sensitive to this choice). The plot shows that the ratio is in steady state, and in the range $30\lesssim\gamma\lesssim 300$,\footnote{This is the energy range between the injection Lorentz factor at $\gamma=3\sigma=30$ and the spectral cutoff at $\gamma\sim 300$, see the inset of \figg{spect_2d}.} it can be fit as 
\begin{equation}\label{eq:eq2}
\frac{dN_{\rm free}/d\gamma}{dN_{\rm trap}/d\gamma}\simeq 0.005\, \gamma
\end{equation}
We have computed the advection time $t_{\rm adv}$ considering the lifetime of particles that remain always trapped in plasmoids. We find that $t_{\rm adv}\sim 0.4 L_x/c$ independently of the Lorentz factor. In retrospect, this is not surprising. The mass-weighted bulk motions of relativistic reconnection are trans-relativistic, with typical outflow velocities of $\sim 0.6 \,c$ \citep{sironi_beloborodov_20}. On average, a trapped particle travels a distance $\sim 0.25 L_x$ before advecting out of the system, which indeed leads to an advection time $t_{\rm adv}\sim 0.4 L_x/c$.  

It follows that the ratio of acceleration time to escape time is energy-independent (for $30\lesssim\gamma\lesssim 300$) and equal to $t_{\rm acc}/t_{\rm esc}\simeq 1.6$, where we have used that $L_x/\comp\sim 1600$ in our reference simulation. This allows to compute the solution of \eq{fp}. As discussed by e.g., \citet{kirk_98}, the solution of \eq{fp} in the case that both the acceleration time and the escape time scale linearly with $\gamma$ is a power law
\begin{equation}
f=\frac{dN_{\rm free}}{d\gamma}\propto \gamma^{-t_{\rm acc}/t_{\rm esc}}
\end{equation}
which for our case yields $dN_{\rm free}/d\gamma\propto \gamma^{-1.6}$, in good agreement with the spectrum measured directly from our simulation.

Based on this model for the acceleration of free particles, one can address the question of what is the power-law slope expected in the asymptotic (and astrophysically-relevant) regime $L_x\gg \comp$. Let us call $s=-d\log N_{\rm free}/d\log \gamma$ the power-law slope of free particles in the asymptotic limit $L_x\gg \comp$. Given that $s=t_{\rm acc}/t_{\rm esc}$ should be independent of the box size $L_x$, this requires that $t_{\rm esc}$ in \eq{eq1} be independent of $L_x$. In turn, given that $t_{\rm adv}\propto L_x/c$, the ratio in \eq{eq2} should scale as $\propto 1/L_x$. In other words, for $d N_{\rm free}/d\gamma=C_{\rm free} \gamma^{-s}$ and $d N_{\rm trap}/d\gamma=C_{\rm trap} \gamma^{-s-1}$, we require $C_{\rm free}/C_{\rm trap}\propto 1/L_x$. 

Let us now consider the specific case of $s=1$, and assume that the spectrum of free particles extends from $\gamma_{\rm min,free}\sim 3\sigma$ up to $\gamma_{\rm cut}\propto L_x$, while the spectrum of trapped particles extends from  $\gamma_{\rm min,free}\sim \sigma$ up to the same $\gamma_{\rm cut}\propto L_x$. The ratio of number of free particles $N_{\rm free}$ to number of trapped particles $N_{\rm trap}$ (which we called $N_{\rm rr}$ for ``reconnection region'' in the main paper) is, in the limit $\gamma_{\rm cut}\gg\gamma_{\rm min,free}\gtrsim\gamma_{\rm min,trap}$,
\be
\frac{N_{\rm free}}{N_{\rm trap}}=\frac{C_{\rm free}}{C_{\rm trap}} \,{\gamma_{\rm min,trap}} \,{\log\left({\frac{\gamma_{\rm cut}}{\gamma_{\rm min,free}}}\right)}
\ee
which, aside from logarithmic corrections, scales as $\propto C_{\rm free}/C_{\rm trap}\propto 1/L_x$. On the other hand, the energy fraction can be written as
\be
\frac{E_{\rm free}}{E_{\rm trap}}=\frac{C_{\rm free}}{C_{\rm trap}}\frac{\gamma_{\rm cut}}{\log({\gamma_{\rm cut}}/{\gamma_{\rm min,trap}})}~,
\ee
which, aside from logarithmic corrections, scales as $\propto(C_{\rm free}/C_{\rm trap})\,\gamma_{\rm cut}\propto {\rm const}$, in agreement with our results in \figg{scan_box} (there, we called $E_{\rm rr}$ the energy content of trapped particles). Based on our model of acceleration, and requiring that the energy fraction of free particles stays constant with box size, we then expect that the spectrum of free particles in the limit $L_x\gg \comp$ should reach an asymptotic power-law slope $s\simeq 1$.

\bibliographystyle{aasjournal}
\bibliography{blob}

\begin{thebibliography}{}
\expandafter\ifx\csname natexlab\endcsname\relax\def\natexlab#1{#1}\fi
\providecommand{\url}[1]{\href{#1}{#1}}
\providecommand{\dodoi}[1]{doi:~\href{http://doi.org/#1}{\nolinkurl{#1}}}
\providecommand{\doeprint}[1]{\href{http://ascl.net/#1}{\nolinkurl{http://ascl.net/#1}}}
\providecommand{\doarXiv}[1]{\href{https://arxiv.org/abs/#1}{\nolinkurl{https://arxiv.org/abs/#1}}}

\bibitem[{{Begelman}(1998)}]{begelman_98}
{Begelman}, M.~C. 1998, \apj, 493, 291, \dodoi{10.1086/305119}

\bibitem[{{B{\"o}ttcher}(2019)}]{bottcher_19}
{B{\"o}ttcher}, M. 2019, Galaxies, 7, 20, \dodoi{10.3390/galaxies7010020}

\bibitem[{{Buneman}(1993)}]{buneman_93}
{Buneman}, O. 1993, {in ``Computer Space Plasma Physics'', Terra Scientific,
  Tokyo, 67}

\bibitem[{{Cerutti} {et~al.}(2020){Cerutti}, {Philippov}, \&
  {Dubus}}]{cerutti_20}
{Cerutti}, B., {Philippov}, A.~A., \& {Dubus}, G. 2020, \aap, 642, A204,
  \dodoi{10.1051/0004-6361/202038618}

\bibitem[{{Cerutti} {et~al.}(2013){Cerutti}, {Werner}, {Uzdensky}, \&
  {Begelman}}]{cerutti_13a}
{Cerutti}, B., {Werner}, G.~R., {Uzdensky}, D.~A., \& {Begelman}, M.~C. 2013,
  \apj, 770, 147, \dodoi{10.1088/0004-637X/770/2/147}

\bibitem[{{Dahlin} {et~al.}(2017){Dahlin}, {Drake}, \& {Swisdak}}]{dahlin_2017}
{Dahlin}, J.~T., {Drake}, J.~F., \& {Swisdak}, M. 2017, Physics of Plasmas, 24,
  092110, \dodoi{10.1063/1.4986211}

\bibitem[{{Daughton} {et~al.}(2011){Daughton}, {Roytershteyn}, {Karimabadi},
  {Yin}, {Albright}, {Bergen}, \& {Bowers}}]{daughton_11}
{Daughton}, W., {Roytershteyn}, V., {Karimabadi}, H., {et~al.} 2011, Nature
  Physics, 7, 539, \dodoi{10.1038/nphys1965}

\bibitem[{{de Gouveia dal Pino} \& {Lazarian}(2005)}]{degouveia_05}
{de Gouveia dal Pino}, E.~M., \& {Lazarian}, A. 2005, \aap, 441, 845,
  \dodoi{10.1051/0004-6361:20042590}

\bibitem[{{de Jager} \& {Harding}(1992)}]{dejager_harding_92}
{de Jager}, O.~C., \& {Harding}, A.~K. 1992, \apj, 396, 161,
  \dodoi{10.1086/171706}

\bibitem[{{Drenkhahn}(2002)}]{drenkhahn_02b}
{Drenkhahn}, G. 2002, \aap, 387, 714, \dodoi{10.1051/0004-6361:20020390}

\bibitem[{{Drenkhahn} \& {Spruit}(2002)}]{drenkhahn_02a}
{Drenkhahn}, G., \& {Spruit}, H.~C. 2002, \aap, 391, 1141,
  \dodoi{10.1051/0004-6361:20020839}

\bibitem[{{Giannios}(2010)}]{giannios_10}
{Giannios}, D. 2010, \mnras, 408, L46, \dodoi{10.1111/j.1745-3933.2010.00925.x}

\bibitem[{{Giannios} \& {Spruit}(2006)}]{giannios_spruit_06}
{Giannios}, D., \& {Spruit}, H.~C. 2006, \aap, 450, 887,
  \dodoi{10.1051/0004-6361:20054107}

\bibitem[{{Giannios} \& {Uzdensky}(2019)}]{giannios_uzdensky_19}
{Giannios}, D., \& {Uzdensky}, D.~A. 2019, \mnras, 484, 1378,
  \dodoi{10.1093/mnras/stz082}

\bibitem[{{Guo} {et~al.}(2014){Guo}, {Li}, {Daughton}, \& {Liu}}]{guo_14}
{Guo}, F., {Li}, H., {Daughton}, W., \& {Liu}, Y.-H. 2014, Physical Review
  Letters, 113, 155005, \dodoi{10.1103/PhysRevLett.113.155005}

\bibitem[{{Guo} {et~al.}(2020){Guo}, {Li}, {Daughton}, {Li}, {Kilian}, {Liu},
  {Zhang}, \& {Zhang}}]{guo_20_b}
{Guo}, F., {Li}, X., {Daughton}, W., {et~al.} 2020, arXiv e-prints,
  arXiv:2008.02743.
\newblock \doarXiv{2008.02743}

\bibitem[{{Hakobyan} {et~al.}(2020){Hakobyan}, {Petropoulou}, {Spitkovsky}, \&
  {Sironi}}]{hakobyan_20}
{Hakobyan}, H., {Petropoulou}, M., {Spitkovsky}, A., \& {Sironi}, L. 2020,
  arXiv e-prints, arXiv:2006.12530.
\newblock \doarXiv{2006.12530}

\bibitem[{{Halzen} \& {Hooper}(2002)}]{halzen_02}
{Halzen}, F., \& {Hooper}, D. 2002, Reports on Progress in Physics, 65, 1025,
  \dodoi{10.1088/0034-4885/65/7/201}

\bibitem[{{Kagan} {et~al.}(2013){Kagan}, {Milosavljevi{\'c}}, \&
  {Spitkovsky}}]{kagan_13}
{Kagan}, D., {Milosavljevi{\'c}}, M., \& {Spitkovsky}, A. 2013, \apj, 774, 41,
  \dodoi{10.1088/0004-637X/774/1/41}

\bibitem[{{Kirk} {et~al.}(1998){Kirk}, {Rieger}, \& {Mastichiadis}}]{kirk_98}
{Kirk}, J.~G., {Rieger}, F.~M., \& {Mastichiadis}, A. 1998, \aap, 333, 452.
\newblock \doarXiv{astro-ph/9801265}

\bibitem[{{Kirk} \& {Skj{\ae}raasen}(2003)}]{kirk_sk_03}
{Kirk}, J.~G., \& {Skj{\ae}raasen}, O. 2003, \apj, 591, 366,
  \dodoi{10.1086/375215}

\bibitem[{{Li} {et~al.}(2019){Li}, {Guo}, {Li}, {Stanier}, \&
  {Kilian}}]{li_2019}
{Li}, X., {Guo}, F., {Li}, H., {Stanier}, A., \& {Kilian}, P. 2019, \apj, 884,
  118, \dodoi{10.3847/1538-4357/ab4268}

\bibitem[{{Lyubarsky} \& {Kirk}(2001)}]{lyubarsky_kirk_01}
{Lyubarsky}, Y., \& {Kirk}, J.~G. 2001, \apj, 547, 437, \dodoi{10.1086/318354}

\bibitem[{{Lyubarsky}(2005)}]{lyubarsky_05}
{Lyubarsky}, Y.~E. 2005, \mnras, 358, 113,
  \dodoi{10.1111/j.1365-2966.2005.08767.x}

\bibitem[{{Lyutikov} \& {Blandford}(2003)}]{lyutikov_03}
{Lyutikov}, M., \& {Blandford}, R. 2003, ArXiv:astro-ph/0312347

\bibitem[{{Milgrom} \& {Usov}(1995)}]{milgrom_1995}
{Milgrom}, M., \& {Usov}, V. 1995, \apjl, 449, L37, \dodoi{10.1086/309633}

\bibitem[{{Petropoulou} \& {Sironi}(2018)}]{petropoulou_18}
{Petropoulou}, M., \& {Sironi}, L. 2018, \mnras, 481, 5687,
  \dodoi{10.1093/mnras/sty2702}

\bibitem[{{Romanova} \& {Lovelace}(1992)}]{romanova_92}
{Romanova}, M.~M., \& {Lovelace}, R.~V.~E. 1992, \aap, 262, 26

\bibitem[{{Sironi} \& {Beloborodov}(2020)}]{sironi_beloborodov_20}
{Sironi}, L., \& {Beloborodov}, A.~M. 2020, \apj, 899, 52,
  \dodoi{10.3847/1538-4357/aba622}

\bibitem[{{Sironi} {et~al.}(2016){Sironi}, {Giannios}, \&
  {Petropoulou}}]{sironi_16}
{Sironi}, L., {Giannios}, D., \& {Petropoulou}, M. 2016, \mnras, 462, 48,
  \dodoi{10.1093/mnras/stw1620}

\bibitem[{{Sironi} \& {Spitkovsky}(2014)}]{ss_14}
{Sironi}, L., \& {Spitkovsky}, A. 2014, \apjl, 783, L21,
  \dodoi{10.1088/2041-8205/783/1/L21}

\bibitem[{{Speiser}(1965)}]{speiser_65}
{Speiser}, T.~W. 1965, \jgr, 70, 4219, \dodoi{10.1029/JZ070i017p04219}

\bibitem[{{Spitkovsky}(2005)}]{spitkovsky_05}
{Spitkovsky}, A. 2005, in AIP Conf. Ser., Vol. 801, Astrophysical Sources of
  High Energy Particles and Radiation, ed. {T.~Bulik, B.~Rudak, \&
  G.~Madejski}, 345

\bibitem[{{Spruit} {et~al.}(2001){Spruit}, {Daigne}, \&
  {Drenkhahn}}]{spruit_01}
{Spruit}, H.~C., {Daigne}, F., \& {Drenkhahn}, G. 2001, \aap, 369, 694,
  \dodoi{10.1051/0004-6361:20010131}

\bibitem[{{Uzdensky}(2020)}]{uzdensky_20}
{Uzdensky}, D.~A. 2020, arXiv e-prints, arXiv:2007.09533.
\newblock \doarXiv{2007.09533}

\bibitem[{{Waxman}(1995)}]{waxman_1995}
{Waxman}, E. 1995, \prl, 75, 386, \dodoi{10.1103/PhysRevLett.75.386}

\bibitem[{{Werner} \& {Uzdensky}(2017)}]{werner_17}
{Werner}, G.~R., \& {Uzdensky}, D.~A. 2017, \apjl, 843, L27,
  \dodoi{10.3847/2041-8213/aa7892}

\bibitem[{{Zenitani} \& {Hoshino}(2007)}]{zenitani_07}
{Zenitani}, S., \& {Hoshino}, M. 2007, \apj, 670, 702, \dodoi{10.1086/522226}

\bibitem[{{Zenitani} \& {Hoshino}(2008)}]{zenitani_08}
---. 2008, \apj, 677, 530, \dodoi{10.1086/528708}

\end{thebibliography}

\end{document}